\documentclass[aps,twocolumn,superscriptaddress]{revtex4}

\usepackage{graphicx}
\usepackage{amsfonts}
\usepackage{amssymb}
\usepackage{amsbsy}
\usepackage{amsmath}
\usepackage{mathrsfs}
\usepackage{latexsym}
\usepackage{natbib}
\usepackage{bm}
\usepackage{subfigure}   
\usepackage{color}
\usepackage{wasysym}
\usepackage{mathbbol}
\usepackage{bigints}
\usepackage{soul}
\allowdisplaybreaks

\def\rs{r_s}
\def\rq{r_Q}
\def\rp{r_{+}}
\def\rm{r_{-}}
\def\kp{\kappa_{+}}
\def\km{\kappa_{-}}

\def\tc{t_c}

\def\rstar{r_{\star}}

\def\rn{Reissner-Nordstr\"om~}

\begin{document}

\title{Optimization of entanglement depends on whether a black hole is extremal}

\author{Subhajit Barman}
\email{subhajit.barman@physics.iitm.ac.in}

\affiliation{Centre for Strings, Gravitation and Cosmology, Department of Physics, Indian Institute of Technology Madras, Chennai 600036, India}

\affiliation{Department of Physics, Indian Institute of Technology Guwahati,
Guwahati 781039, Assam, India}



\author{Bibhas Ranjan Majhi}
\email{bibhas.majhi@iitg.ac.in}

\affiliation{Department of Physics, Indian Institute of Technology Guwahati,
Guwahati 781039, Assam, India}


\date{\today}

\begin{abstract}

We consider two Unruh-DeWitt detectors interacting with a massless, minimally coupled scalar field in a $(1+1)$ dimensional Reissner-Nordstr\"om black hole spacetime. In particular, one of the detectors, corresponding to \emph{Alice}, is moving along an outgoing null trajectory. While the other detector carried by \emph{Bob} is static. With this set-up, we investigate the entangling condition and the measure of the entanglement, concurrence, in the nonextremal and extremal scenarios. Our observations suggest, as expected, a qualitative similarity in characteristics of the entanglement between these two scenarios.
However, we find quantitative differences between the nonextremal and extremal concurrences for a broad range of black hole charges.
With moderately large detector transition energy, the extremal background always accounts for the larger entanglement than the nonextremal one. In contrast, with low detector transition energy, entanglement on the nonextremal background can be greater. Therefore, by adjusting the detector transition energy, one can perceive optimum entanglement from either the extremal or the nonextremal background.

\end{abstract}

\maketitle

\section{Introduction}\label{Introduction}

The phenomenon of quantum entanglement has characteristics regarded to have a pure quantum mechanical origin. These quantum characteristics have been verified through the experimental observations of the violation of the Bell inequality, which the classical theory of local hidden variables could not explain. Apart from laying the foundation of quantum mechanics, quantum entanglement also presents some very useful and conducive applications. One of the forerunners of these applications is the possibility of communication without eavesdropping, which is prioritized in quantum communication and cryptography \cite{Tittel:1998ja, Salart-2008}. Another exciting prediction of entanglement with many possible future applications is quantum teleportation \cite{Hotta:2008uk, Hotta:2009, Frey:2014}.

In the last few decades, entanglement has also become an important arena for studying the dynamics of relativistic quantum particles, especially through particle detectors. Some of the directions that are considered in these study, broadly comprising the area of relativistic quantum information (RQI) \cite{Reznik:2002fz, Lin:2010zzb, Ball:2005xa, Cliche:2009fma, MartinMartinez:2012sg, Cai:2018xuo, Zhou:2017axh, Floreanini:2004, Pan:2020tzf, Barman:2022xht}, include understanding the radiative process of entangled relativistic particles \cite{Menezes:2015veo, Menezes:2015iva, Rodriguez-Camargo:2016fbq, Costa:2020aqa, Cai:2019pnw, Liu:2018zod, Cai:2017jan, Barman:2021oum, Barman:2022utm}, studying entanglement dynamics \cite{Menezes:2015uaa, Menezes:2017rby, Menezes:2017oeb}, studying entangled Unruh Otto engines \cite{Kane:2021rhg, Barman:2021igh}, and entanglement swapping from background field to quantum probes \cite{VALENTINI1991321, Reznik:2002fz, Reznik:2003mnx}. In particular, the phenomenon of entanglement swapping, also known as entanglement harvesting, is becoming increasingly interesting due to the possibility of the entanglement being used for further quantum information purposes \cite{Koga:2018the}. In the simplest setup, it essentially deals with the possibility of uncorrelated systems getting entangled over time due to their background geometry, motion, and many other key factors. There is an enormous number of works arriving presently that actively pursue understanding these entanglement profiles for Unruh-DeWitt detectors perturbatively interacting with the background field. These works range from inertial detectors in flat spacetime \cite{Koga:2018the, Barman:2022xht} to detectors in different trajectories in curved spacetime \cite{MartinMartinez:2012sg, Gallock-Yoshimura:2021yok, K:2023oon}. Many of these works also study the effects of acceleration \cite{Reznik:2002fz, Floreanini:2004, Koga:2019fqh}, circular motion \cite{Zhang:2020xvo}, thermal bath \cite{Brown:2013kia, Barman:2021bbw}, and also the passing of gravitational wave \cite{Cliche:2010fi, Xu:2020pbj, Barman:2023aqk}. 
Whereas, in \cite{Chowdhury:2021ieg, Barman:2022loh} the authors study the possibility of entanglement enhancement or degradation starting with two correlated Unruh-DeWitt detectors.
Other detector models \cite{Brown:2012pw, Brown:2013kia, Makarov:2017erw} that interact non-perturbatively with the background field also present possibilities of entanglement utilizing them.

In our current work, we intend to understand the above-mentioned entanglement profiles of Unruh-DeWitt detectors in a $(1+1)$ dimensional charged Reissner-Nordstr\"om black hole spacetime.
It is known that the Reissner-Nordstr\"om black holes exhibit extremality, a characteristic trait of stationary black holes. Very few works in the literature discuss entanglement dynamics in observationally relevant stationary black hole backgrounds. The computational complexity has contributed to this lack of investigation. However, one can study the effects of some scenarios particular to these stationary black holes, like extremality, present in some other computationally simpler backgrounds, like the RN black holes. This is one of our motivations for considering this particular background. Furthermore, it is accepted in literature \cite{Vanzo:1995bh, Angheben:2005rm, Vanzo:2011wq}, that in the extremal limit, i.e., when the two horizons of a charged black hole merge to a single one, the Hawking emission ceases to exist.
This phenomenon is visualized by taking the extremal limit from the nonextremal Hawking spectra \cite{Liberati:2000sq, Barman:2018ina}, and also starting from an extremal black hole altogether \cite{Gao:2002kz, Balbinot:2007kr, Barman:2018ina, Ghosh:2021ijv}. 
On the other hand, entanglement depends on both local terms given by the individual detectors transition probabilities and non-local term indicating the correlation between two different detectors through the background field.
Due to these local terms, in the extremal scenario, one may expect to have different entanglement properties than the nonextremal case. This also motivates one to study the entanglement in the extremal limit of a \rn background and understand the differences in comparison to the nonextremal scenario. However, from the study of the area law of black hole entropy, it is encountered in the literature that the result of an extremal black hole from the beginning and the limit from nonextremal to extremal do not coincide \cite{Preskill:1991tb, Hawking:1995fd, Ghosh:1996gp}. In such cases, it is often suggested to start with an extremal black hole from the beginning to get the extremal results \cite{Ghosh:1996gp, Gao:2002kz, Pradhan:2012yx}. Therefore, in this work, we consider the extremal and nonextremal \rn black hole backgrounds separately from the beginning and investigate the entanglement profiles in them.
In this direction, the simplest set-up for the detectors will be two static ones in the \rn background. 
%
However, with two static detectors, one may not be able to get any entanglement from the Boulware vacuum as these detectors see this vacuum as no different than the Minkowski vacuum, and static detectors do not give entanglement from the Minkowski vacuum with infinite switching \cite{Koga:2018the}.
%
Therefore, in the next plausible detector set-up, one can consider one detector to be static and another in some sort of motion. We consider this other detector in outgoing null trajectory. In particular, we consider that one of the detectors, denoted by $A$ corresponding to \emph{Alice}, is moving along an outgoing null trajectory and the other detector $B$ carried by \emph{Bob} to be static. Furthermore, with this detector set-up, we study the entanglement from the Boulware-like vacuum in the \rn black hole background. We will find that such a choice of vacuum and trajectories nullifies the effects of particle production on the entanglement. The choice of $(1+1)$-dimensional background provides flexibility to perform the computations analytically. Just to mention we avoid extending our analysis to any other type of vacuum, namely the Unruh or Hartle-Hawking-like, as defining a Kruskal-like coordinate transformation in the extremal scenario is not a straightforward task.

Our observations suggest a qualitative similarity in the entanglement profiles between the nonextremal and extremal scenarios. In both cases, entanglement monotonically decreases with increasing detector transition energy. This is probably due to the fact that the nature of spacetime does not change drastically outside the horizon. However, there are some quantitative differences. For instance, for low detector transition energy, one may get more entanglement from the nonextremal background. While for moderately high transition energy, entanglement from the extremal background is always the maximum. The entanglement in both instances is periodic with respect to the distance $d_{n}$ that distinguishes different null paths and with respect to the distance $d_{s}$ of the static detector. With increasing detector transition energy, this periodicity increases, and the amplitude decreases.

We would also like to emphasize that one could have considered physically more relevant quantum states, such as the thermal states. However, before investigating the intricacies involved with a thermal state it is imperative to understand the situation without any thermal fluctuations. It should also be noted that the consideration of a thermal background will also make the computations much more involved. On the other hand, one could have also considered the set-up in a more relevant higher dimensions. However, in this regard, we would like to point out that as long as physics near the event horizon is concerned the field theory is analogous to a $(1+1)$ dimensional scenario. Other the other hand, this $(1+1)$ dimensional \rn metric can be considered a solution of $(1+1)$ dimensional Dilaton theory (see pages 21-22 of \cite{Grumiller:2002nm}), which is a dimensionally reduced theory of higher dimensional Einstein-Hilbert action. Moreover, as we have considered static and radial null trajectories for the detectors, the information about the specific angular coordinates become redundant, and thus one can treat this problem in an $(1+1)$ dimensional set-up. In this respect, analysis in $(1+1)$ dimensions can be relevant for understanding higher dimensional physics.

On the other hand, we would like to point out that if one considers timelike trajectories for one or two of the detectors the computations become much more complicated. The analytical calculations may not remain possible any more, and one may have to resort to numerical computations. In the present work the choice of null paths makes the computations much simpler and analytically traceable. This analytical computation is important as it provides a much needed understanding as to why the extremal and nonextremal black hole backgrounds should be treated separately. Additionally to get a preliminary idea about the different responses of extremal and non-extremal backgrounds on the quantum effects, it is important to start with a simple scenario. As mentioned above, the null path provides the same. Moreover, from the theoretical point of view the consideration of null trajectories can be of interest as hypothetical detectors such as massless qubit detectors can be utilized in these scenarios. One the other hand, one can expect that detectors that closely resemble these null paths, such as detectors with very high velocity, can provide similar outcomes.

This article is organized in the following manner. In Sec. \ref{sec:model-set-up}, we provide a model set-up for understanding the entangling condition and the quantification of the entanglement with two Uruh-DeWitt detectors. In Sec. \ref{sec:RN-BH-spacetime} we consider the $(1+1)$ dimensional \rn black hole spacetime and understand the condition for extremality. In the following Sec. \ref{sec:det-paths-GreensFn}, we elaborate on the detector trajectories and construct the necessary Green's functions. Subsequently, in Sec. \ref{sec:entanglement-harvesting}, we study individual detector transition probabilities, the entanglement condition, and the concurrence in the nonextremal and extremal scenarios. We conclude this work with a discussion and the implications of our findings in Sec. \ref{sec:discussion}.

\section{Model set-up}\label{sec:model-set-up}

We begin our study in this section with a model setup of entanglement. In particular, we follow the work \cite{Koga:2018the} of recent times, where the authors considered a proper time ordering to construct the necessary Green's functions. Similar approaches were also perceived in \cite{Koga:2019fqh}.

We consider two point-like two-level Unruh-DeWitt detectors $A$ and $B$, associated with two distinct observers, Alice and Bob, respectively. For the $j^{th}$ detector, the energy eigenstates are considered to be $|E_{n}^{j}\rangle$, where $n\,(=0,1)$ signify different energy levels. These states are, in general considered non-degenerate, i.e., $E_{1}^{j}\neq E_{0}^{j}$, $\Delta E^{j} = E_{1}^{j}- E_{0}^{j}>0$ gives the transition energy. 
%
We consider these detectors interacting with a massless, minimally coupled background real scalar field $\Phi(X)$, with $X$ denoting the spacetime point. The corresponding interaction action is
\begin{eqnarray}\label{eq:TimeEvolution-int}
 S_{int} &=& \int_{-\infty}^{\infty} c\,\bigg[ 
\kappa_{A}(\tau_{A})m^{A}(\tau_{A}) \Phi\left(X_{ A}(\tau_{A})\right) 
d\tau_{A} \nonumber\\ 
&& +~ \kappa_{B}(\tau_{B})m^{B}(\tau_{B}) \Phi\left(X_{ 
B}(\tau_{B})\right) d\tau_{B}\bigg]~.
\end{eqnarray}
Here, $c$, $\kappa_{j}(\tau_{j})$, $m^{j}(\tau_{j})$, and $\tau_{j}$ respectively denote the couplings between the individual detectors and the scalar field, the switching functions, the monopole moment operators, and the individual detector proper times. At the same time, $X_{j}(\tau_{j})$ denotes the individual detector world lines. In the asymptotic past the initial detector field state is assumed to be $|in\rangle = |0\rangle |E_{0}^{A}\rangle |E_{0}^{B}\rangle$, where $|0\rangle$ represents the field's ground state. With time evolution one obtains the final detector field state in asymptotic future as $|out\rangle = \mathcal{T}\left\{e^{i S_{int}}|in\rangle\right\}$, where $\mathcal{T}$ denotes time ordering. 
%
We treat the coupling strengths $c$ perturbatively to get the explicit expression of this state. After tracing out the field degrees of freedom, in the basis of the detector states $\big\{|E_{1}^{A}\rangle |E_{1}^{B}\rangle, |E_{1}^{A}\rangle |E_{0}^{B}\rangle, |E_{0}^{A}\rangle |E_{1}^{B}\rangle, |E_{0}^{A}\rangle |E_{0}^{B}\rangle\big\}$, one can obtain the reduced detector density matrix as given in Eq. (2.2) of \cite{Barman:2021bbw},
%
\begin{equation}\label{eq:detector-density-matrix}
 \rho_{AB} = 
 {\left[\begin{matrix}
 0 & 0 & 0 & c^2\,\varepsilon\\~\\
0 & c^2\,P_{A} & c^2\,P_{AB} & 0\\~\\
0 & c^2\,P_{AB}^{*} & c^2\,P_{B} & 0\\~\\
c^2\,\varepsilon^{*} & 0 & 
0 & 
1-c^2\,P_{A}-c^2\,P_{B}
 \end{matrix}\right]}
 +\mathcal{O}(c^4)~.
\end{equation}
%
In that density matrix the quantities that will be relevant for our current goal are $P_{j}$ and $\varepsilon$, which have explicit expressions 
\begin{subequations}
\begin{eqnarray}
P_{j} &=& |\langle E^{j}_{1}|m^{j}(0)|E^{j}_{0}\rangle|^2\,\mathcal{I}_{j}~,\\
\varepsilon &=& \langle E^{A}_{1}|m^{A}(0)|E^{A}_{0}\rangle\,\langle E^{B}_{1}|m^{B}(0)|E^{B}_{0}\rangle\,\mathcal{I}_{\varepsilon}~,
\end{eqnarray}
\end{subequations}
%
where the quantities $\mathcal{I}_{j}$ and $\mathcal{I}_{\varepsilon}$ are given by 
\begin{eqnarray}
 \mathcal{I}_{j} &=& \int_{-\infty}^{\infty}d\tau'_{j} 
\int_{-\infty}^{\infty}d\tau_{j}~e^{-i\Delta E^{j}(\tau'_{j}-\tau_{j})} 
G_{W}(X'_{j},X_{j}),\nonumber
\end{eqnarray}
\begin{eqnarray}\label{eq:all-integrals}
\mathcal{I}_{\varepsilon} &=& -i\int_{-\infty}^{\infty}d\tau'_{B} 
\int_{-\infty}^{\infty}d\tau_{A}~\scalebox{0.91}{$e^{i(\Delta 
E^{B}\tau'_{B}+\Delta E^{A}\tau_{A})} G_{F}(X'_{B},X_{A}).$}\nonumber\\
\end{eqnarray}
In our current work we have considered the detectors eternally interacting with the field, i.e., $\kappa_{j}(\tau_{j})=1$, which also reflects in the expressions of the previous equations. Furthermore, we recognize the functions $G_{W}(X_{j},X_{j'})$, $G_{F}(X_{j},X_{j'})$, and $G_{R}(X_{j},X_{j'})$ as the the positive frequency Wightman function, the Feynman propagator, and the retarded Green's function respectively. The positive frequency Wightman function is associated with $X_{j}>X_{j'}$, where $j'\neq j$. These Green's functions, see \cite{Koga:2018the}, are defined as
%
\begin{eqnarray}\label{eq:Greens-fn-gen}
 G_{W}\left(X_{j},X_{j'}\right) &\equiv& \langle 
0|\Phi\left(X_{j}\right)\Phi\left(X_{j'}\right)|0
\rangle~,\nonumber\\
 G_{F}\left(X_{j},X_{j'}\right) &\equiv& -i\langle 
0|\mathcal{T}\left\{\Phi\left(X_{j}\right)\Phi\left(X_{j'}\right)\right\}|0
\rangle~,\nonumber\\
 G_{R}\left(X_{j},X_{j'}\right) &\equiv& i\theta(t-t')\langle 
0|\left[\Phi\left(X_{j'}\right),\Phi\left(X_{j}\right)\right]|0
\rangle,\nonumber\\
\end{eqnarray}
where, $|0 \rangle$ denotes the field's ground state. 
%
For bipartite systems \cite{Peres:1996dw, Horodecki:1996nc} it is observed that the entanglement is possible only when the partial transposition of the reduced detector density matrix has negative eigenvalue. Considering the reduced density matrix as obtained in our case (see Eq. (2.2) of article \cite{Barman:2021bbw}), this condition is fulfilled only when
\begin{equation}\label{eq:EH-cond1}
 P_{A}P_{B}<|\varepsilon|^2~. 
\end{equation}
In terms of the integrals from Eq. (\ref{eq:all-integrals}) this condition becomes \cite{Koga:2018the, Koga:2019fqh}
\begin{equation}\label{eq:cond-entanglement}
 \mathcal{I}_{A}\mathcal{I}_{B}<|\mathcal{I}_{\varepsilon}|^2~.
\end{equation}
This condition is obtained considering a perturbation up-to the order of $c^2_{j}$. We shall be using this expression with the considered perturbative order in $c_{j}$ to study the entanglement phenomenon, like done in \cite{Koga:2018the}.

Now following the procedure of \cite{Koga:2018the}, one can represent the Feynman propagator in terms of the positive frequency Wightman functions as $iG_{F}\left(X_{j},X_{j'}\right) = G_{W}\left(X_{j},X_{j'}\right) + i G_{R}\left(X_{j'},X_{j}\right) = G_{W}\left(X_{j},X_{j'}\right) + \theta(T'-T) \left\{G_{W} \left(X_{j'},X_{j}\right)-G_{W}\left(X_{j},X_{j'}\right)\right\}$, where $T$ and $T'$ are time coordinates with respect to which the positive frequency field modes are defined. We use this expression to simplify the expression of the integral $\mathcal{I}_{\varepsilon}$ from Eq. (\ref{eq:all-integrals}) as
%
\begin{eqnarray}\label{eq:Ie-integral}
 && \mathcal{I}_{\varepsilon} = -\int_{-\infty}^{\infty}d\tau_{B} 
\int_{-\infty}^{\infty}d\tau_{A}~\scalebox{0.91}{$e^{i(\Delta 
E^{B}\tau_{B}+\Delta E^{A}\tau_{A})} $}\times\nonumber\\
~&& \scalebox{0.87}{$\left[G_{W}(X_{B},X_{A})+ \theta(T_{A}-T_{B}) 
\left\{G_{W}\left(X_{A},X_{B}\right)-G_{W}\left(X_{B},X_{A}\right)\right\}
\right].$}\nonumber\\
\end{eqnarray}
Thus all the integrals $\mathcal{I}_{A}$, $\mathcal{I}_{B}$  and $\mathcal{I}_{\varepsilon}$ for estimating the entanglement condition (\ref{eq:cond-entanglement}) are now expressed in terms of the Wightman functions.

There are different measures like negativity and concurrence to quantify the entanglement \cite{Zyczkowski:1998yd, Vidal:2002zz, Eisert:1998pz, Devetak_2005}, once the condition for entanglement (\ref{eq:cond-entanglement}) is satisfied. For instance, negativity, given by the sum of all negative eigenvalues of the partial transpose of $\rho_{AB}$, corresponds to the upper bound of the distillable entanglement. Whereas, concurrence $\mathcal{C}(\rho_{AB})$ is relevant for obtaining the entanglement of formation \cite{Bennett:1996gf, Hill:1997pfa, Wootters:1997id, Koga:2018the, Koga:2019fqh}. The concurrence \cite{Koga:2018the, Koga:2019fqh, Hu:2015lda} is the most frequently used measure, which for two qubits system \cite{Koga:2018the} is given by
\begin{eqnarray}\label{eq:concurrence-gen-exp}
 \mathcal{C}(\rho_{AB}) &=& 
max\bigg[0,~ 2c^2 
\left(|\varepsilon|-\sqrt{P_{A}P_{B}}\right)+\mathcal{O}(c^4)\bigg]\nonumber\\
~&\approx& max\bigg[0,~ 2c^2|\langle E_{1}^{B}|m_{B}(0)| E_{0}^{B}\rangle| |\langle 
E_{1}^{A}|m_{A}(0)| E_{0}^{A}\rangle|\nonumber\\
~&& ~~~~~~\times 
\left(|\mathcal{I}_{\varepsilon}|-\sqrt{\mathcal{I}_{A}\mathcal{I}_{B}}
\right)\bigg]~,
\end{eqnarray}
Here we have considered both detectors interacting with the background field with the same strength, i.e., $c_{A}=c_{B}=c$. The multiplicative factors $|\langle E_{1}^{j}|m_{j}(0)| E_{0}^{j}\rangle|$ are reminiscent of the internal structure of the detectors. Therefore, in order to understand the effects of the motion of the detectors and background spacetime in entanglement, it is convenient to study
\begin{equation}\label{eq:concurrence-I}
\mathcal{C}_{\mathcal{I}} = \left(|\mathcal{I} 
_{\varepsilon}| -\sqrt{\mathcal{I}_{A}\mathcal{I}_{B}} \right)~.
\end{equation} 
In fact, we shall endeavor to estimate this quantity to understand the entanglement in nonextremal and extremal Reissner-Nordstr\"om black hole spacetime.

\section{The Reissner-Nordstr\"om black hole spacetime}\label{sec:RN-BH-spacetime}

We now consider a $(1+1)$ dimensional \rn black hole spacetime. The $(3+1)$ dimensional \rn black hole spacetime is a solution of the Einstein-Maxwell equation, and due to spherical symmetry of this solution one can obtain its $(1+1)$ dimensional form by dropping the angular components, see \cite{Juarez-Aubry:2015dla, Juarez-Aubry:2021tae}. Our reason behind considering $(1+1)$ dimensions is that in this dimensions the spacetime is conformally flat. In particular, in $(1+1)$ dimensions the \rn metric looks like
\begin{eqnarray}\label{eq:metric-RN}
 ds^2 &=& -\Big(1-\frac{\rs}{r}+\frac{\rq^2}{r^2}\Big)\,d\tc^2+ 
\Big(1-\frac{\rs}{r}+\frac{\rq^2}{r^2}\Big)^{-1} dr^2 ,\nonumber\\
\end{eqnarray}
where, $\rs=2G\,M$ correspond to the mass and $\rq=G^{1/2}\,Q$ correspond to the charge of the black hole, with $G$ being the Newton's gravitational constant. One should note that here we have considered the black hole to be nonextremal in general, i.e., $\rq\neq \rs/2$. The extremal limit is given by $\rq\to \rs/2$. In this black hole spacetime the two horizons are respectively located at
\begin{eqnarray}\label{eq:expression-rpm}
 r_{\pm} = \frac{1}{2}\,\Big(\rs\pm\sqrt{\rs^2-4\rq^2}\Big)~,
\end{eqnarray}
where $r_{+}$ represents the outer event horizon and $r_{-}$ the inner Cauchy horizon. The surface gravities of these horizons are $\kappa_{\pm} = (r_{+}-r_{-})/
2r_{\pm}^2$. In the extremal limit these two horizons merge together at $\rs/2$ and the surface gravity vanishes.

One can obtain the tortoise coordinate $\rstar$, in a nonextremal Reissner-Nordstr\"om black hole spacetime from 
\begin{equation}\label{eq:tortoise-coord}
 d\rstar = \frac{dr}{1-\rs/r+\rq^2/r^2}~.
\end{equation}
After integrating with a suitable choice of integration constant this tortoise coordinate becomes \cite{Liberati:2000sq, Gao:2002kz, Balbinot:2007kr}
\begin{equation}\label{eq:tortoise-NE}
 \rstar = r+\frac{1}{2\kp} \ln \Big(\frac{r}{\rp}-1\Big)-\frac{1}{2\km} \ln \Big(\frac{r}{\rm}-1\Big)~.
\end{equation}
While the tortoise coordinate in an extremal Reissner-Nordstr\"om black hole spacetime is obtained from $d\rstar = dr/(1-\rs/2r)^2$. In this case the functional form of $\rstar$, see \cite{Liberati:2000sq, Gao:2002kz, Balbinot:2007kr, Barman:2018ina}, is 
\begin{equation}\label{eq:tortoise-E}
 \rstar = r+\rs \ln \Big(\frac{r}{\rs/2}-1\Big)-\frac{(\rs/2)^2}{r-\rs/2}~.
\end{equation}

In terms of the tortoise coordinate (\ref{eq:tortoise-coord}) the previous $(1+1)$ dimensional Reissner-Nordstr\"om metric Eq. (\ref{eq:metric-RN}) is given by
\begin{equation}\label{eq:metric-tortoise-RN}
 ds^2 = \Big(1-\frac{\rs}{r}+\frac{\rq^2}{r^2}\Big)\big[-d\tc^2+ 
d\rstar^2\big]~.
\end{equation}
This expression denotes a conformally flat spacetime with the conformal factor $(1-\rs/r+\rq^2/r^2)$. One can now decompose the field operator in terms of the modes expressed in terms of $\tc$ and $\rstar$. The annihilation operator in the field operator annihilates the conformal vacuum, which is the Boulware vacuum.

We should mention that as pointed out in \cite{Gao:2002kz}, one cannot get the expression of the extremal tortoise coordinate (\ref{eq:tortoise-E}) just by putting $\rq=\rs/2$ in the nonextremal one (\ref{eq:tortoise-NE}). In that case one usually encounters a zero by zero situation. Therefore, it is encouraged to consider the nonextremal and extremal scenarios separately from the beginning \cite{Gao:2002kz, Ghosh:1996gp}. However, we have observed that by taking a limit $\rq\to\rs/2$ in the nonextremal tortoise coordinate, one can obtain the extremal result (\ref{eq:tortoise-E}), see Appendix \ref{Appn:tortoise-NE-to-E}. The zero by zero situation is resolved as one takes this limit in the level of tortoise coordinate. In the same section of the Appendix, we explain why it is necessary to consider the nonextremal and extremal scenarios separately from the beginning when estimating integrals $\mathcal{I}_{j}$ and $\mathcal{I}_{\varepsilon}$.

\section{Detector trajectories and Wightman functions}\label{sec:det-paths-GreensFn}

\subsection{Null paths related to particle creation from the conformal vacuum}

In this section we are going to consider specific world lines for the observers. In particular, we consider \emph{Alice}, denoted by detector $A$, to be in an outgoing null trajectory. While \emph{Bob}, denoted by detector $B$, remains static outside of the \rn event horizon. For an observer along null trajectory either of the coordinates $u=\tc-\rstar$ and $v=\tc+\rstar$ is fixed. For instance, along outgoing null trajectory $u$ is fixed, while along ingoing null trajectory $v$ is fixed. These coordinates are sometimes also referred as the retarded and the advanced time coordinates. For an observer in outgoing null trajectory, it is convenient to define the Eddington-Finkelstein (EF) coordinates $(t,r)$ to observe particle creation from the Boulware like vacuum. These coordinates are defined as $t+r=\tc+\rstar$, and in terms of them the conformal \rn metric from Eq. (\ref{eq:metric-tortoise-RN}) becomes
\begin{eqnarray}\label{eq:metric-RN-EF}
 ds^2 &=& -\Big(1-\frac{\rs}{r}+\frac{\rq^2}{r^2}\Big)\, dt^2 + 2\,\Big(\frac{\rs}{r}-\frac{\rq^2}{r^2}\Big)\, dtdr\nonumber\\
 ~&& ~~~~~+ ~
\Big(1+\frac{\rs}{r}-\frac{\rq^2}{r^2}\Big)\, dr^2~.
\end{eqnarray}
In terms of the EF coordinates the outgoing null trajectory is found as the positive solution of the equation $ds^2=0$, which gives

\begin{equation}\label{eq:outgoing-null-path}
 \frac{dt}{dr} = \frac{1+\rs/r-\rq^2/r^2}{1-\rs/r+\rq^2/r^2}=f(r)~,
\end{equation}
which further provides the path as
\begin{equation}\label{eq:outgoing-null-coord}
 t = -r+2\,\rstar+d_{n}~.
\end{equation}
Here $d_{n}$ is the integration constant, which signifies the distance between the different outgoing null paths. One can notice that this definition of an outgoing null path in terms of the EF coordinates is true in both the nonextremal and extremal cases. Then in the nonextremal case one should use the definition of the tortoise coordinate as given by Eq. (\ref{eq:tortoise-NE}). While in the extremal case one should use Eq. (\ref{eq:tortoise-E}).

On the other hand for a static detector $(\tc-t)$ is constant. We consider this constant to be $c_{1}$, so that $\tc=t+c_{1}$, and $dt=d\tc$. For Bob's detector denoted by $B$ we shall be using this trajectory.

Here, one should note that it is not possible to define a proper time for an observer along the null trajectory.
In this regard, one usually considers an affine parameter with respect to which one can talk about the four momenta of the observer \cite{Chakraborty:2019ltu}. This affine parameter corresponding to the null trajectory can be identified as the Eddington-Finkelstein (EF) time ($t_{j}$) that we have used in our case. 
In \cite{Dalui:2020qpt}, for a general spherically symmetric static black hole, it has been explicitly shown that for the particular null trajectories EF time serves as one of the affine parameters. 
In this regard, one can look at Appendix \ref{Appn:Affine-null} for an elaborate discussion.

\subsection{Green's function corresponding to the two detectors}\label{Sec:Greens-fn-null-paths}

In terms of the retarded and advanced time coordinates the positive frequency Boulware modes are $e^{-i\omega u}$ and $e^{-i\omega v}$. We decompose a massless minimally coupled scalar field $\Phi$ using these modes and introducing the sets of ladder operators $\{\hat{a}_{k}^{B\dagger}, \hat{a}_{k}^{B}\}$ and $\{\hat{b}_{k}^{B\dagger}, \hat{b}_{k}^{B}\}$ as \cite{Hodgkinson:2013tsa}
\begin{eqnarray}
 \Phi &=& \int_{0}^{\infty} \frac{d\omega_{k}}{\sqrt{4\pi\omega_{k}}} 
\Big[\hat{a}_{k}^{B} e^{-i\omega_{k} u} + \hat{a}_{k}^{B\dagger} 
e^{i\omega_{k} u} \nonumber\\
~&& ~~~~~~~~+~~ \hat{b}_{k}^{B} e^{-i\omega_{k} v} + 
\hat{b}_{k}^{B\dagger} 
e^{i\omega_{k} v} \Big]~.
\end{eqnarray}
Here non-vanishing commutators between the annihilation and creation operators are $\big[ \hat{a}_{k}^{B}, \hat{a}_{k'}^{B\dagger}\big]=\delta_{k,k'}$ and $\big[ \hat{b}_{k}^{B}, \hat{b}_{k'}^{B\dagger}\big]=\delta_{k,k'}$. The annihilation operators annihilate the Boulware vacuum $|0\rangle_{B}$, i.e., $\hat{a}_{k}^{B} |0\rangle_{B} = 0 = \hat{b}_{k}^{B} |0\rangle_{B}$. Using this field decomposition and the commutation relations one can get the positive frequency Wightman function as
\begin{eqnarray}\label{eq:Wightman-Boulware}
 G^{+}_{B}(X_{j},X_{l}) &=& _{B}\langle 0|\Phi(X_{j})\Phi(X_{l}) 
|0\rangle_{B}\nonumber\\
~&=& \int_{0}^{\infty} \frac{d\omega_{k}}{4\pi\omega_{k}} \big[e^{-i\omega_{k} 
(u_{j}-u_{l})} + e^{-i\omega_{k} 
(v_{j}-v_{l})}\big]~. \nonumber\\
\end{eqnarray}
The subscript $j$ and $l$ respectively correspond to the $j^{th}$ and $l^{th}$ detectors, which relate to the events $X_{j}$ and $X_{l}$.

As we have previously mentioned the detector $A$ moves along a null path. Then using Eq. (\ref{eq:outgoing-null-coord}) we have $t_{A} = -r_{A}+2\,r_{\star_{A}}+d_{n}$. Furthermore using this Eq. (\ref{eq:outgoing-null-coord}) one has the quantities
\begin{eqnarray}\label{eq:vv-rel-A}
 v'_{A}-v_{A} &=& t'_{c_{A}}+r'_{\star_{A}}-(t_{c_{A}}+r_{\star_{A}})\nonumber\\
 ~&=& 2(r'_{\star_{A}}-r_{\star_{A}})~,
\end{eqnarray}
and
\begin{eqnarray}\label{eq:uu-rel-A}
 u'_{A}-u_{A} &=& t'_{c_{A}}-r'_{\star_{A}}-(t_{c_{A}}-r_{\star_{A}}) \nonumber\\
 ~&=& 0~.
\end{eqnarray}
The detector $B$ is static at some radial distance, and we consider $r'_{\star_{B}} = r_{\star_{B}} =d_{s}$. Then for detector $B$ we have 
\begin{eqnarray}\label{eq:vu-rel-B}
 v'_{B}-v_{B} = u'_{B}-u_{B}  = t'_{c_{B}}-t_{c_{B}} =  t'_{B}-t_{B}
\end{eqnarray}
Using trajectory for detector $A$ from Eq. (\ref{eq:outgoing-null-coord}) and the trajectory for the detector $B$, one can also obtain the quantities
\begin{eqnarray}\label{eq:vv-rel-AB}
 v_{A}-v_{B} &=& t_{c_{A}}+r_{\star_{A}}-(t_{c_{B}}+r_{\star_{B}})\nonumber\\
 ~&=& 2\,r_{\star_{A}}+d_{n}-(t_{c_{B}}+d_{s})~,
\end{eqnarray}
and
\begin{eqnarray}\label{eq:uu-rel-AB}
 u_{A}-u_{B} &=& t_{c_{A}}-r_{\star_{A}}-(t_{c_{B}}-r_{\star_{B}}) \nonumber\\
 ~&=& d_{n}-(t_{c_{B}}-d_{s})~.
\end{eqnarray}
Furthermore, whenever we are talking about the detector $A$ the expression of the tortoise coordinate should be taken from Eq. (\ref{eq:tortoise-NE}) or (\ref{eq:tortoise-E}), depending on whether we are considering a nonextremal or extremal black hole spacetime. On the other hand, when one considers the static detector $B$ the relation between the coordinate and the EF time is $t_{c_{B}}=t_{B}+c_{1}$.

\section{Entanglement}\label{sec:entanglement-harvesting}

In this section we study the entanglement condition in the nonextremal and extremal Reissner-Nordstr\"om black hole background, with one of the Unruh-DeWitt detectors static and another moving in an outgoing null trajectory. In particular, we shall begin our study by evaluating the integrals $\mathcal{I}_{j}$ and $\mathcal{I}_{\varepsilon}$ in the nonextremal and extremal scenarios respectively. Then we will estimate the individual detector transition probabilities and the concurrence in both the cases and compare the results.

We would also like to mention that in our expressions of the local and non-local terms in entanglement from Eq. \eqref{eq:all-integrals} the integrals are defined with respect to the individual proper times of the detectors. However, as mentioned before it is not possible to obtain a proper time for detectors that follow null-like trajectories. To circumvent this issue we have considered a suitable affine parameter, which happens to be the Eddington-Finkelstein time $t_{j}$ (see Appendix \ref{Appn:Affine-null}), with respect to which we will have to define all our integrals. This basically implies that for the considered null trajectories the initial action from Eq. \eqref{eq:TimeEvolution-int} is now better described by the affine parameter $t_{j}$. Thus later on the evolution operator and consequent integrals of Eq. \eqref{eq:all-integrals} will also be described by this affine parameter. 

It should also be noted that this choice of the affine parameter is not unique and $t_{j}$ multiplied by any function of the spacetime point that remains constant on the trajectory will also serve as an affine parameter. However, upto our knowledge $t_{j}$ is the simplest of such describing outgoing principle null congruences. At the same time, as long as our calculation is concerned those different affine parameters will have little effect on the final results as we will represent our integrals in terms of the radial coordinate $r$, which is integrated over the region outside of the horizon to infinity. To elucidate this let us focus on the expression of the EF time $t_{j}$ from Eq. \eqref{eq:outgoing-null-coord}. One can notice that even if the right side of that expression is multiplied by a constant factor, the characteristics of $t_{j}$ remains the same as one approaches $r\to\rp$ or $r\to \infty$.

\subsection{Nonextremal scenario}\label{subsec:entanglement-harvesting-NE}

\subsubsection{Evaluation of the integral $\mathcal{I}_{j}$}

First, we consider evaluating the integral $\mathcal{I}_{j}$ in a nonextremal \rn background. This particular quantity signifies individual detector transition probability and acts as a local contribution in the concurrence. Using the expression of the Green's function from Eq. (\ref{eq:Wightman-Boulware}) one can write the integral $\mathcal{I}_{j}$ from Eq. (\ref{eq:all-integrals}) as 
\begin{eqnarray}\label{eq:Ij-RN-NE-1}
 \mathcal{I}_{j} &=& \int_{-\infty}^{\infty} dt'_{j}\int_{-\infty}^{\infty} 
dt_{j}~ e^{-i\Delta E^{j}(t'_{j}-t_{j})} G^{+}_{B}(X'_{j},X_{j})\nonumber\\
~&=& \int_{0}^{\infty} d\omega_{k}~ 
\mathcal{I}_{j_{\omega_{k}}}~,
\end{eqnarray}
where the integral $\mathcal{I}_{j_{\omega_{k}}}$ in general, i.e., in both the nonextremal and extremal cases, looks like 
\begin{eqnarray}\label{eq:Ij-RN-NE-2}
 \mathcal{I}_{j_{\omega_{k}}} &=& \frac{1}{4\pi\omega_{k}}\int_{-\infty}^{\infty} 
dt'_{j}\int_{-\infty}^{\infty} 
dt_{j}\, e^{-i\Delta E^{j}(t'_{j}-t_{j})}\nonumber\\
~&&~~ \times~\big[e^{-i\omega_{k} 
(u_{j'}-u_{j})} + e^{-i\omega_{k} 
(v_{j'}-v_{j})}\big]~.
\end{eqnarray}
One should note that this integral $\mathcal{I}_{j_{\omega_{k}}}$ now signifies individual detector transition probability corresponding to a certain field mode frequency $\omega_{k}$, see \cite{Scully:2017utk, Kolekar:2013hra, Barman:2021oum, Barman:2021kwg}. It is often convenient to define the necessary quantities corresponding to fixed $\omega_{k}$, as done in \cite{Barman:2021kwg}, and here also we shall follow the same procedure.

In $(1+1)$ dimensions one can obtain the dimension of the coupling strength $c_{A}=c_{B}=c$ to be $[L^{-1}]$ in natural units, which can be understood from the interaction action of Eq. (\ref{eq:TimeEvolution-int}) as the scalar field $\Phi$ is here dimensionless. One should note here that $[L]$ corresponds to the dimension of length. Again the integral $\mathcal{I}_{j}$ multiplied with $c^2$ and $|\langle E^{j}_{1}|m_{j}(0)|E^{j}_{0}\rangle|^2$ (dimensionless) provides the actual dimensionless transition probability. Therefore, the integral $\mathcal{I}_{j}$ has a dimension of $[L^2]$. One should note that $c$ is a constant parameter and does not depend on the detector trajectory or the background spacetime. Thus $\mathcal{I}_{j}$ itself signifies the transition probability, atleast the part of the transition probability that depends on different system variables nontrivially. Furthermore, in a similar sense, since $\mathcal{I}_{j_{\omega_k}}$ integrated over $\omega_k$ gives $\mathcal{I}_{j}$, $\mathcal{I}_{j_{\omega_k}}$ is expected to correspond to transition probability for a specific field mode frequency, and has the dimension of $[L^3]$.

\vspace{0.2cm}
\emph{\underline{Detector $A$}:-} Now let us concentrate on a certain detector trajectory. We consider Alice's detector, which is in an outgoing null path. For detector $A$ using Eqs. (\ref{eq:vv-rel-A}) and (\ref{eq:uu-rel-A}) we can express the previous integral of $\mathcal{I}_{j_{\omega_{k}}}$ as
\begin{eqnarray}\label{eq:IA-RN-NE-1}
 \mathcal{I}_{A_{\omega_{k}}} &=& \frac{1}{4\pi\omega_{k}}\int_{-\infty}^{\infty} 
dt'_{A}\int_{-\infty}^{\infty} 
dt_{A}\, e^{-i\Delta E\,(t'_{A}-t_{A})}\nonumber\\
~&&~~ \times~\big[1 + e^{-2i\omega_{k} 
(r'_{\star_{A}}-r_{\star_{A}})}\big]~.
\end{eqnarray}
Due to the integration of the first term in the bracket there will be a square of Dirac delta distribution $\delta(\Delta E)$, which will eventually vanish for our considered $\Delta E>0$. Then using Eq. (\ref{eq:outgoing-null-path}) and for a nonextremal black hole (Eq. (\ref{eq:tortoise-NE})) we get 
\begin{eqnarray}\label{eq:IA-RN-NE-2}
 \mathcal{I}_{A_{\omega_{k}}} &=& \frac{1}{4\pi\omega_{k}}\int_{\rp}^{\infty} 
f(r'_{A})dr'_{A}\int_{\rp}^{\infty} 
f(r_{A})dr_{A}\, \nonumber\\
~&& \times~ e^{-i(\Delta E+2\omega_{k})(r'_{A}-r_{A})}~\bigg(\tfrac{r'_{A}-\rp}{r_{A}-\rp}\bigg)^{-i(\Delta 
E + \omega_{k})/\kp}\,\nonumber\\
~&& \times\bigg(\tfrac{r'_{A}-\rm}{r_{A}-\rm}\bigg)^{i(\Delta 
E + \omega_{k})/\km}~.
\end{eqnarray}
Now let us consider change of variables $y'_{A}=r'_{A}/\rp-1$ and $y_{A}=r_{A}/\rp-1$. With this the integral simplifies to 
\begin{eqnarray}\label{eq:IA-RN-NE-3}
 \mathcal{I}_{A_{\omega_{k}}} &=& \frac{\rp^2}{4\pi\omega_{k}} \Bigg|\int_{0}^{\infty} dy_{A}\, 
\frac{2(y_{A}+1)^2\rp}{y_{A}\big[(y_{A}+1)\rp-\rm\big]}\nonumber\\
~&& \,\frac{e^{i\rp(\Delta E+2\omega_{k})y_{A}}}{y_{A}^{-\frac{i(\Delta 
E+\omega_{k})}{\kp}}}\, \bigg[\frac{(y_{A}+1)\rp}{\rm}-1\bigg]^{-\frac{i(\Delta 
E+\omega_{k})}{\km}}\Bigg|^2.\nonumber\\
\end{eqnarray}
The integrand in the above expression is oscillatory in nature, which when integrated over an infinite range does not converge. One can circumvent this issue by introducing a regulator of the form of $e^{-\epsilon\, y_{A}}\,y_{A}^{\epsilon}$ in the integrand, where $\epsilon$ is very small positive real parameter. In the limit $\epsilon\to 0$ the regulator $e^{-\epsilon\, y_{A}}\,y_{A}^{\epsilon}$ becomes a multiplicative factor of unity and thus the exact expression of the above integral is obtained as one takes $\epsilon\to 0$ in the final result. For similar regularization techniques one can look into \cite{Barman:2018ina}.
An enthusiastic reader may go through Appendix \ref{Appn:Ij-NE} for an explicit analytic expression of this integral.

Here we shall like to point out that we could have performed the integration over the first term in Eq. (\ref{eq:IA-RN-NE-1}) by considering a change of variables $t'_{A}-t_{A}=\Bar{u}_{A}$ and $t'_{A}+t_{A}=\Bar{v}_{A}$. In that scenario one would have obtained an outcome of $\delta(0)\,\delta(\Delta E)$, which has diverging $\delta(0)$ in it, and making the first term vanish does not remain that straight forward. A suitable way out of this situation is to define time rate of transition probability, which for the infinite switching scenario is obtained by dividing the entire $\mathcal{I}_{A_{\omega_{k}}}$ by $\delta(0)$. Interesting to notice that this rate vanishes due to the presence of $\delta(\Delta E)$ in it.

Note that the $(1+1)$ dimensional Wightman function of Eq. (\ref{eq:Wightman-Boulware}) contained both the infrared and ultraviolet divergences, which becomes evident as $\omega_k\to 0$ and $\omega_k\to\infty$. We have not explicitly evaluated this Wightman function and kept it in a mode-sum form to be integrated in the response function (\ref{eq:Ij-RN-NE-1}). We are going to define all our required quantities corresponding to a certain field mode frequency like done in Eq. (\ref{eq:Ij-RN-NE-1}). Thus we avoid performing the integrations over $\omega_k$, and also avoid encountering the infrared or ultraviolet divergences. Therefore, in this sense, our results are valid for a particular mode frequency.
\vspace{0.2cm}

\emph{\underline{Detector $B$}:-} Here we consider Bob's detector, which is kept static outside of the black hole event horizon. For detector $B$ we use the coordinate relations from Eq. (\ref{eq:vu-rel-B}) and the integral $\mathcal{I}_{B_{\omega_{k}}}$ becomes
\begin{eqnarray}\label{eq:IB-RN-NE-1}
 \mathcal{I}_{B_{\omega_{k}}} &=& \frac{2}{4\pi\omega_{k}} \int_{-\infty}^{\infty} 
dt'_{B}\int_{-\infty}^{\infty} 
dt_{B}\, e^{-i(\Delta E+\omega_{k})\,(t'_{B}-t_{B})}~.\nonumber\\
\end{eqnarray}
Naturally this integral will generate a factor of square of $\delta(\omega_{k}+\Delta E)$. From Eq. (\ref{eq:Ij-RN-NE-1}) we recall that there was a integration over $\omega_{k}$ from zero to infinity, and we have considered $\Delta E>0$. Then this Dirac delta distribution will have vanishing contribution in the integration range of $\omega_{k}$. Therefore, $\mathcal{I}_{B_{\omega_{k}}}$, where $B$ corresponds to a static detector, will vanish, i.e., $\mathcal{I}_{B_{\omega_{k}}} = 0$.

Here we should mention that one could also consider a change of variables of $t'_{B}-t_{B}=\Bar{u}_{B}$ and $t'_{B}+t_{B}=\Bar{v}_{B}$ to evaluate the previous integral of Eq. (\ref{eq:IB-RN-NE-1}). In that case there would have been a factor of $\delta(0)\, \delta(\omega_{k}+\Delta E)$ in the outcome, which contains a divergent term $\delta(0)$, and the previous result of $\mathcal{I}_{B_{\omega_{k}}} = 0$ seems not to be completely true. However, here one can circumvent these issues to obtain a consistent result by defining time rates. For infinite switching this rate is taken by excluding the factor of $\delta(0)$ from $\mathcal{I}_{B_{\omega_{k}}}$, and we define this rate to be $\Bar{\mathcal{I}}_{B_{\omega_{k}}}$. One can observe that it contains a factor of $\delta(\omega_{k}+\Delta E)$ and thus vanishes. In the evaluation of the nonlocal entangling term $\mathcal{I}_{\varepsilon}$ also we shall observe the occurrence of the $\delta(0)$, and shall suitably define the time rates to avoid these seeming divergences.
\vspace{0.2cm}

\subsubsection{Evaluation of the integral $\mathcal{I}_{\varepsilon}$}
Second, we consider evaluating the integral $\mathcal{I}_{\varepsilon}$, which contains the signatures of both the detectors. This quantity acts as a nonlocal contribution to the concurrence. With the motivation to evaluate this integral $\mathcal{I}_{\varepsilon}$ from Eq. (\ref{eq:all-integrals}), we first express it as
\begin{equation}\label{eq:Ie-RN-NE-1}
 \mathcal{I}_{\varepsilon} = -\mathcal{I}^{W}_{\varepsilon} - 
\mathcal{I}^{R}_{\varepsilon}~,
\end{equation}
where each of the quantities $\mathcal{I}^{W}_{\varepsilon}$ and $\mathcal{I}^{R}_{\varepsilon}$ are expressed in a similar fashion like Eq. (\ref{eq:Ij-RN-NE-1}), as
\begin{eqnarray}\label{eq:IeW-RN-NE-1}
 \mathcal{I}^{W}_{\varepsilon} &=& \int_{-\infty}^{\infty}dt_{A} 
\int_{-\infty}^{\infty}dt_{B}~e^{i\Delta 
E(t_{B}+t_{A})} G_{W}(X_{B},X_{A})\nonumber\\
~&=& \int_{0}^{\infty} d\omega_{k}~ 
\mathcal{I}^{W}_{\varepsilon_{\omega_{k}}}~,
\end{eqnarray}
and
\begin{eqnarray}\label{eq:IeR-RN-NE-1}
~\mathcal{I}^{R}_{\varepsilon} &=& \int_{-\infty}^{\infty}dt_{A} 
\int_{-\infty}^{\infty}dt_{B}~e^{i\Delta 
E(t_{B}+t_{A})}  \theta(t_{c_{A}}-t_{c_{B}})\nonumber\\
~&& ~~~~~~~~~\times \big[G_{W}\left(X_{A},X_{B}\right)-G_{W}\left(X_{B},X_{A}
\right) \big]~\nonumber\\
~&=& \int_{0}^{\infty} d\omega_{k}~ 
\mathcal{I}^{R}_{\varepsilon_{\omega_{k}}}~.
\end{eqnarray}
Using the general expression of the Green's function from Eq. (\ref{eq:Wightman-Boulware}) one can obtain the expressions of these integrals $\mathcal{I}^{W}_{\varepsilon_{\omega_{k}}}$ and $\mathcal{I}^{R}_{\varepsilon_{\omega_{k}}}$ given by
\begin{eqnarray}\label{eq:IeW-RN-NE-2}
 \mathcal{I}^{W}_{\varepsilon_{\omega_{k}}} &=& \frac{1}{4\pi\omega_{k}} \int_{-\infty}^{\infty} 
dt_{A}\int_{-\infty}^{\infty} dt_{B}~ e^{i\Delta E(t_{B}+t_{A})}~\nonumber\\
~&& \times~\big[e^{-i\omega_{k} (u_{B}-u_{A})} + e^{-i\omega_{k} (v_{B}-v_{A})}\big]~,
\end{eqnarray}
and 
\begin{eqnarray}\label{eq:IeR-RN-NE-2}
 \mathcal{I}^{R}_{\varepsilon_{\omega_{k}}} &=& \frac{1}{4\pi\omega_{k}} \int_{-\infty}^{\infty} 
dt_{A}\int_{-\infty}^{\infty} dt_{B}\, e^{i\Delta E(t_{B}+t_{A})}\,\theta(t_{c_{A}}-t_{c_{B}})\nonumber\\
~&& \times\big[e^{-i\omega_{k} (u_{A}-u_{B})} + 
e^{-i\omega_{k} 
(v_{A}-v_{B})} \nonumber\\
~&& - e^{-i\omega_{k} (u_{B}-u_{A})} - e^{-i\omega_{k} 
(v_{B}-v_{A})}\big]~.
\end{eqnarray}
One may notice that up-to this point everything holds for both the nonextremal and extremal cases and for arbitrary detector trajectories. In the following investigations we shall explicitly consider the detector trajectories and background.

\vspace{0.2cm}
\underline{\emph{Evaluation of $\mathcal{I}^{W}_{\varepsilon_{\omega_{k}}}$}}:- In order to evaluate $\mathcal{I}^{W}_{\varepsilon_{\omega_{k}}}$ we consider the relations from Eq. (\ref{eq:vv-rel-AB}) and (\ref{eq:uu-rel-AB}) in Eq. (\ref{eq:IeW-RN-NE-2}). This expression now looks like
\begin{eqnarray}\label{eq:IeW-RN-NE-3}
 \mathcal{I}^{W}_{\varepsilon_{\omega_{k}}} &=& \frac{e^{i\omega_{k} \,(d_{n}-c_{1})}}{4\pi\omega_{k}}\,\int_{-\infty}^{\infty} 
dt_{A}\int_{-\infty}^{\infty} dt_{B}~ e^{i\Delta E(t_{B}+t_{A})}~\nonumber\\
~&& \times~\big[e^{-i\omega_{k} (t_{B}-d_{s})} + e^{i\omega_{k} (2\,r_{\star_{A}}-t_{B}-d_{s})}\big]\nonumber\\
~&=& \frac{2\pi}{4\pi\omega_{k}}\,\delta(\omega_{k}-\Delta E)\,e^{i\omega_{k} \,(d_{n}-c_{1})}\,\int_{-\infty}^{\infty} 
dt_{A}\, e^{i\Delta E\,t_{A}}~\nonumber\\
~&& \times~\big[e^{i\omega_{k} \,d_{s}} + e^{i\omega_{k} (2\,r_{\star_{A}}-d_{s})}\big]~.
\end{eqnarray}
Integrating the first term in the bracket will provide a multiplicative factor of $\delta(\Delta E)$, which vanishes for $\Delta E>0$. Therefore, we shall only be concerned about the second term. Let us define $\mathcal{I}^{W}_{\varepsilon_{\omega_{k}}} = \bar{\mathcal{I}}^{W}_{\varepsilon}(\omega_{k})\,\delta[(\omega_{k}-\Delta E)\rs]$. We have included a multiplicative factor of $\rs$ inside the Dirac delta to make its argument dimensionless. Then with a change of variables to $y_{A}$, the integral $\bar{\mathcal{I}}^{W}_{\varepsilon}(\omega_{k})$ can be represented as
\begin{eqnarray}\label{eq:IeW-RN-NE-4}
 \bar{\mathcal{I}}^{W}_{\varepsilon}(\omega_{k}) &=& \frac{\rp^2\rs}{\omega_{k}}\,e^{i\omega_{k} \,(d_{n}-c_{1}-d_{s})}\,e^{i\Delta E \,d_{n}}~\nonumber\\
~&\times&~\int_{0}^{\infty} dy_{A}\, 
\frac{(y_{A}+1)^2}{y_{A}\big[(y_{A}+1)\rp-\rm\big]}\nonumber\\
~&& \tfrac{e^{i\rp(\Delta E+2\omega_{k})(y_{A}+1)}}{y_{A}^{-\frac{i(\Delta 
E+\omega_{k})}{\kp}}}\, \bigg[\tfrac{(y_{A}+1)\rp}{\rm}-1\bigg]^{-\frac{i(\Delta 
E+\omega_{k})}{\km}}~.\nonumber\\
\end{eqnarray}
Similar to Eq. \eqref{eq:IA-RN-NE-3} the above integral is also oscillatory, which does not converge. In a similar manner as mentioned before one can perform the above integration  analytically introducing a regulator of the form $e^{-\epsilon\, y_{A}}\,y_{A}^{\epsilon}$, see Appendix \ref{Appn:Ie-NE}.
We should mention that $\bar{\mathcal{I}}^{W}_{\varepsilon}(\omega_{k})$ is nonzero only when $\omega_{k}=\Delta E$, due to the Dirac delta distribution. We shall then take $\bar{\mathcal{I}}^{W}_{\varepsilon}(\Delta E)$ for our purpose to estimate the concurrence.
\vspace{0.2cm}

\underline{\emph{Evaluation of $\mathcal{I}^{R}_{\varepsilon_{\omega_{k}}}$}}:- Let us now evaluate the integral $\mathcal{I}^{R}_{\varepsilon_{\omega_{k}}}$. We mention that the Heaviside theta function $\theta(t_{c_{A}}-t_{c_{B}})$ form this integral can be removed with the change of integration limit in $t_{c_{B}}$ from $(-\infty,\infty)$ to $(-\infty,t_{c_{A}}]$. Again $t_{c_{A}}$ is given by $t_{c_{A}} = r_{\star_{A}}+d_{n}$. With these transformations the concerned integral now becomes
\begin{eqnarray}\label{eq:IeR-RN-NE-3}
 \mathcal{I}^{R}_{\varepsilon_{\omega_{k}}} &=& \frac{1}{4\pi\omega_{k}} \int_{-\infty}^{\infty} 
dt_{A}\int_{-\infty}^{r_{\star_{A}}+d_{n}} dt_{c_B}\, e^{i\Delta E(t_{c_B}-c_{1}+t_{A})}\,\nonumber\\
~&& \times\big[e^{-i\omega_{k} (d_{n}-t_{c_{B}}+d_{s})} + 
e^{-i\omega_{k} 
(2\,r_{\star_{A}}+d_{n}-t_{c_{B}}-d_{s})} \nonumber\\
~&& - e^{i\omega_{k} (d_{n}-t_{c_{B}}+d_{s})} - 
e^{i\omega_{k} 
(2\,r_{\star_{A}}+d_{n}-t_{c_{B}}-d_{s})} \big]~.
\end{eqnarray}
To perform this integral we introduce regulator of the form $e^{\bar{\epsilon}\, t_{c_{B}}}$. The outcome is given by
\begin{eqnarray}\label{eq:IeR-RN-NE-4}
 \mathcal{I}^{R}_{\varepsilon_{\omega_{k}}} &=& -\frac{i}{4\pi\omega_{k}} \int_{-\infty}^{\infty} 
dt_{A} \,e^{i\Delta E(r_{\star_{A}}+d_{n}-c_{1}+t_{A})}\,\nonumber\\
~&& \times ~e^{\bar{\epsilon}(r_{\star_{A}}+d_{n})} \,\big[e^{i\omega_{k} (d_{s}-r_{\star_{A}})} + 
e^{-i\omega_{k} (d_{s}-r_{\star_{A}})} \big]\nonumber\\
~&& \times~\Big[\frac{1}{\omega_{k}+\Delta E-i\bar{\epsilon}}+\frac{1}{\omega_{k}-\Delta E+i\bar{\epsilon}}\Big]~.
\end{eqnarray}
Now let us see how we represent this integral in a fashion similar to the expression of the integral $\mathcal{I}^{W}_{\varepsilon_{\omega_{k}}}$. According to the Sokhotski-Plemelj theorem \cite{book:Birrell} one can express 
\begin{eqnarray}\label{eq:SP-theorem}
\lim_{\bar{\epsilon}\to0+}~\frac{1}{z\mp i\bar{\epsilon}} =\pm i \pi \,\delta(z) +\mathcal{P}\big(1/z\big)~,
\end{eqnarray}
where, $\mathcal{P}\big(1/z\big)$ corresponds to the principal value of $(1/z)$, and is a finite quantity. Then it is evident that only the second term inside the bracket in Eq. (\ref{eq:IeR-RN-NE-4}), which now has a multiplicative factor of $\delta(\omega_{k}-\Delta E)$, will contribute to the integral compared to other terms. Let us explain it in a further simplified form by taking a common factor of $\delta(\omega_{k}-\Delta E)$ out of this integral. Then all the terms with principle values and $\delta(\omega_{k}+\Delta E)$ in the numerator will have $\delta(\omega_{k}-\Delta E)$ in their denominator, and those terms vanish for $\omega_{k}=\Delta E$, see \cite{Barman:2021bbw}. These perceptions motivate us to define $\mathcal{I}^{R}_{\varepsilon_{\omega_{k}}} = \bar{\mathcal{I}}^{R}_{\varepsilon}(\omega_{k})\,\delta[(\omega_{k}-\Delta E)\rs]$, where the contributing term in $\bar{\mathcal{I}}^{R}_{\varepsilon}(\omega_{k})$ is given by
\begin{eqnarray}\label{eq:IeR-RN-NE-5}
 \bar{\mathcal{I}}^{R}_{\varepsilon}(\omega_{k}) &=& -\frac{\pi\,\rs}{4\pi\omega_{k}} \int_{-\infty}^{\infty} 
dt_{A} \,e^{i\Delta E(r_{\star_{A}}+d_{n}-c_{1}+t_{A})}\,\nonumber\\
~&& \times ~\big[e^{i\omega_{k} (d_{s}-r_{\star_{A}})} + 
e^{-i\omega_{k} (d_{s}-r_{\star_{A}})} \big]~.\nonumber\\
\end{eqnarray}
With a change of variables to $y_{A}$ this previous expression becomes
\begin{eqnarray}\label{eq:IeR-RN-NE-6}
 \bar{\mathcal{I}}^{R}_{\varepsilon}(\omega_{k}) &=& -\frac{\rp^2\rs\,e^{i\Delta E(d_{n}-c_{1})}}{2\,\omega_{k}} \,\int_{0}^{\infty} dy_{A}\, 
\tfrac{(y_{A}+1)^2}{y_{A}\big[(y_{A}+1)\rp-\rm\big]}\nonumber\\
~&\times& e^{2i\rp\Delta E(y_{A}+1)}\,y_{A}^{\frac{3i\Delta 
E}{2\kp}}\, \bigg[\tfrac{(y_{A}+1)\rp}{\rm}-1\bigg]^{-\frac{3i\Delta 
E}{2\km}}\nonumber\\
~&\times& \Bigg\{ e^{i\omega_{k}d_{s}-i\rp\omega_{k}(y_{A}+1)}\,y_{A}^{-\frac{i\omega_{k}}{2\kp}}\, \bigg[\tfrac{(y_{A}+1)\rp}{\rm}-1\bigg]^{\frac{i\omega_{k}}{2\km}}\nonumber\\
~&+& e^{-i\omega_{k}d_{s}+i\rp\omega_{k}(y_{A}+1)}\,y_{A}^{\frac{i\omega_{k}}{2\kp}}\, \bigg[\tfrac{(y_{A}+1)\rp}{\rm}-1\bigg]^{-\frac{i\omega_{k}}{2\km}} \Bigg\}\,.\nonumber\\
\end{eqnarray}
One can also perform this integration analytically introducing a regulator of the form $e^{-\epsilon\, y_{A}}\,y_{A}^{\epsilon}$, see Appendix \ref{Appn:Ie-NE}.

\subsection{Extremal scenario}\label{subsec:entanglement-harvesting-E}

\subsubsection{Evaluation of the integral $\mathcal{I}_{j}$}

For the evaluation of $\mathcal{I}_{j}$ we consider its decomposition as provided in Eq. (\ref{eq:Ij-RN-NE-1}) and (\ref{eq:Ij-RN-NE-2}), and shall actually evaluate $\mathcal{I}_{j_{\omega_{k}}}$, which corresponds to individual detector transition probability for fixed field mode frequency.
\vspace{0.2cm}

\emph{\underline{Detector $A$}:-} In the extremal scenario also when evaluating $\mathcal{I}_{j_{\omega_{k}}}$ for detector $A$, the previous Eq. (\ref{eq:IA-RN-NE-1}) is valid. However now we will have to use the expression of the tortoise coordinate $\rstar$ from Eq. (\ref{eq:tortoise-E}) corresponding to an extremal \rn black hole. In particular, with this substitution the integral $\mathcal{I}_{j_{\omega_{k}}}$ becomes
\begin{eqnarray}\label{eq:IA-RN-E-1}
 \mathcal{I}_{A_{\omega_{k}}} &=& \frac{1}{4\pi\omega_{k}}\,\int_{\rp}^{\infty} 
f(r'_{A})dr'_{A}\int_{\rp}^{\infty} 
f(r_{A})dr_{A}\, \nonumber\\
~&& \times~ e^{-i(\Delta E+2\omega_{k})(r'_{A}-r_{A})}\,\bigg(\tfrac{r'_{A}-\rs/2}{r_{A}-\rs/2}\bigg)^{-2i\rs(\Delta 
E + \omega_{k})}\nonumber\\
~&& \times~ e^{-2i(\Delta E+\omega_{k})(\rs/2)^2\left(\frac{1}{r'_A-\rs/2}-\frac{1}{r_A-\rs/2}\right)}~.
\end{eqnarray}
Now let us consider change of variables $\bar{y}'_{A}=r'_{A}/(\rs/2)-1$ and $\bar{y}_{A}=r_{A}/(\rs/2)-1$. With this the integral simplifies to 
\begin{eqnarray}\label{eq:IA-RN-E-2}
 \mathcal{I}_{A_{\omega_{k}}} &=& \frac{(\rs/2)^2}{4\pi\omega_{k}}\, \Bigg|\int_{0}^{\infty} d\bar{y}_{A}\, 2\, (\bar{y}_{A}+1)^2 ~ \bar{y}_{A}^{-2+2 i \rs (\Delta E+\omega_{k})} \nonumber\\
 ~&\times& \exp \left(\frac{1}{2} i \rs \bar{y}_{A} (\Delta E+2 \omega_{k})-\tfrac{i \rs (\bar{y}_{A}+1) (\Delta E+\omega_{k})}{\bar{y}_{A}}\right)\Bigg|^2.\nonumber\\
\end{eqnarray}
Here also the integral is oscillatory, and thus will not converge. This particular integral can be evaluated introducing a regulator of the form $e^{-\epsilon \,\bar{y}_{A}-\epsilon /\bar{y}_{A}}\,\bar{y}_{A}^{\epsilon}$, the explicit analytic expression of which is provided in Appendix \ref{Appn:Ij-E}. Similar regularization techniques are also adopted in \citep{Barman:2018ina, Ghosh:2021ijv} to handle the divergences therein.
\vspace{0.1cm}

\emph{\underline{Detector $B$}:-} The expression of the integral $\mathcal{I}_{B_{\omega_{k}}}$ even in the extremal case is the same as the nonextremal case from Eq. (\ref{eq:IB-RN-NE-1}). Then here also for a static detector $B$ the integral $\mathcal{I}_{B_{\omega_{k}}} = 0$.
\vspace{0.2cm}

\subsubsection{Evaluation of the integral $\mathcal{I}_{\varepsilon}$}
In extremal case also we use the representations from Eqs. (\ref{eq:Ie-RN-NE-1}), (\ref{eq:IeW-RN-NE-1}), and (\ref{eq:IeR-RN-NE-1}) like the nonextremal case. Furthermore, we shall evaluate the nonlocal terms $\mathcal{I}^{W}_{\varepsilon_{\omega_{k}}}$ and $\mathcal{I}^{R}_{\varepsilon_{\omega_{k}}}$ following the prescriptions from Eqs. (\ref{eq:IeW-RN-NE-2}) and (\ref{eq:IeR-RN-NE-2}).
\vspace{0.2cm}

\underline{\emph{Evaluation of $\mathcal{I}^{W}_{\varepsilon_{\omega_{k}}}$}}:- Upto Eq. (\ref{eq:IeW-RN-NE-3}) from the nonextremal case is valid still in the extremal scenario as we have not yet specified the functional form of the tortoise coordinate. Then using the expression of the tortoise coordinate from Eq. (\ref{eq:tortoise-E}) and with a change of variables to $\bar{y}_{A}$, one can obtain the expression of 
$\bar{\mathcal{I}}^{W}_{\varepsilon}(\omega_{k})$ as
\begin{eqnarray}\label{eq:IeW-RN-E-1}
 \bar{\mathcal{I}}^{W}_{\varepsilon}(\omega_{k}) &=& \frac{\rs^2}{2\,\omega_{k}}\,e^{i\omega_{k} \,(d_{n}-c_{1}-d_{s})}\,e^{i\Delta E \,d_{n}}~\nonumber\\
~&& \times~\int_{0}^{\infty} d\bar{y}_{A}\, 
(\bar{y}_{A}+1)^2 \,\bar{y}_{A}^{-2+2 i \rs (\Delta E+\omega_{k})} \nonumber\\
~&\times& \exp \left(\tfrac{i \rs (\bar{y}_{A}+1) (2 \omega_{k} (\bar{y}_{A}-1)+\Delta E (\bar{y}_{A}-2))}{2 \bar{y}_{A}}\right)~.\nonumber\\
\end{eqnarray}
We introduce regulator of the form $e^{-\epsilon \,\bar{y}_{A}-\epsilon /\bar{y}_{A}}\,\bar{y}_{A}^{\epsilon}$ to evaluate this integral, see Appendix \ref{Appn:Ie-E}. It is to be noted that for our purpose to estimate the concurrence we shall use $\bar{\mathcal{I}}^{W}_{\varepsilon}(\Delta E)$.
\vspace{0.2cm}

\underline{\emph{Evaluation of $\mathcal{I}^{R}_{\varepsilon_{\omega_{k}}}$}}:- The general expressions of the integrals from Eqs. (\ref{eq:IeR-RN-NE-3}), (\ref{eq:IeR-RN-NE-4}), and (\ref{eq:IeR-RN-NE-5}) are still valid in the extremal scenario. In particular, we use the expression (\ref{eq:IeR-RN-NE-5}) with the tortoise coordinate for the extremal black hole (\ref{eq:tortoise-E}) and the change of variables to $\bar{y}_{A}$ to evaluate $\bar{\mathcal{I}}^{R}_{\varepsilon}(\omega_{k})$. This integral now takes the form
\begin{widetext}
\begin{eqnarray}\label{eq:IeR-RN-E-1}
 \bar{\mathcal{I}}^{R}_{\varepsilon}(\omega_{k}) &=& -\frac{\rs^2}{4\,\omega_{k}}\,e^{i\Delta E(d_{n}-c_{1})}\int_{0}^{\infty} d\bar{y}_{A}\, 
\, (\bar{y}_{A}+1)^2 \,\bar{y}_{A}^{-2}~\bigg\{\bar{y}_{A}^{-i \rs (\omega_{k}-3 \Delta E)} ~\exp \left(i \,d_{s} \,\omega_{k}-\tfrac{i \rs (\bar{y}_{A}+1) (\omega_{k} (\bar{y}_{A}-1)+\Delta E (3-2 \bar{y}_{A}))}{2 \bar{y}_{A}}\right) \nonumber\\
~ && +~\bar{y}_{A}^{i \rs (\omega_{k}+3 \Delta E)} ~\exp \left(-i \,d_{s} \,\omega_{k}+\tfrac{i \rs (\bar{y}_{A}+1) (\omega_{k} (\bar{y}_{A}-1)-\Delta E (3-2 \bar{y}_{A}))}{2 \bar{y}_{A}}\right)~\bigg\}\,.
\end{eqnarray}
\end{widetext}
One can perform this integration analytically by introducing a regulator of the form $e^{-\epsilon \,\bar{y}_{A}-\epsilon /\bar{y}_{A}}\,\bar{y}_{A}^{\epsilon}$, see Appendix \ref{Appn:Ie-E}. In the subsequent part we shall use the explicit expressions of $\bar{\mathcal{I}}^{W}_{\varepsilon}(\Delta E)$ and $\bar{\mathcal{I}}^{R}_{\varepsilon}(\Delta E)$ to estimate the concurrence in the nonextremal and extremal scenarios.

\begin{figure*}
\centering
\includegraphics[width=0.47\linewidth]{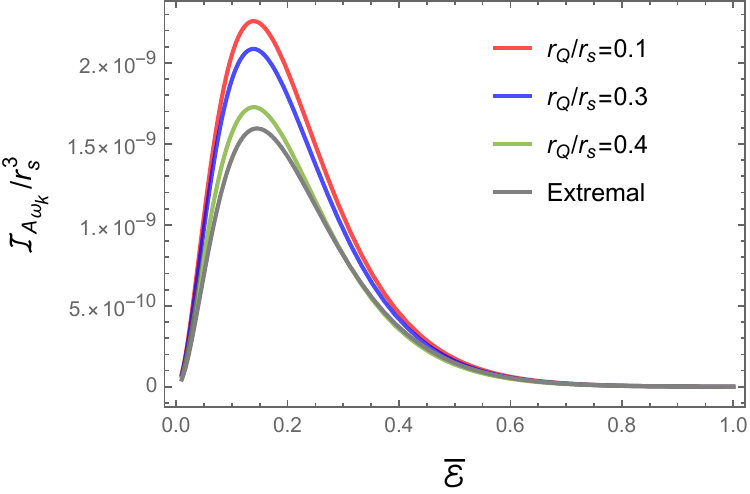}
\hskip 10pt
\includegraphics[width=0.47\linewidth]{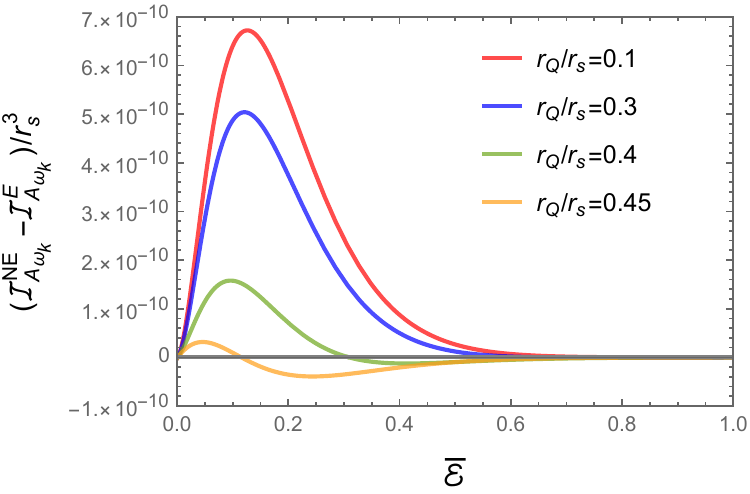}
\caption{In left we have plotted the individual detector transition probabilities $\mathcal{I}_{A_{\omega_{k}}}$ of detector $A$ as a function of the dimensionless energy gap $\bar{\mathcal{E}}=\rs\,\Delta E$ in both the nonextremal and extremal scenarios. To obtain these plots we have fixed the dimensionless frequency of the field at $\bar{\omega}_{k}=\rs\,\omega_{k}=1$. In right we have plotted the difference between the nonextremal and extremal $\mathcal{I}_{A_{\omega_{k}}}$ as functions of $\bar{\mathcal{E}}$ for different $\rq/\rs$. One observes that nonextremal detector transition is always greater than extremal, when $\rq/\rs$ is low. However, near extremal limit $(\rq/\rs\to 0.5)$, when $\rq/\rs= 0.4$ the nonextremal transition becomes lower compared to the extremal case for moderately large $\bar{\mathcal{E}}$. This feature is more prominent in $\rq/\rs> 0.4$ regime.}\label{fig:Ij-dE-NE-E}
\end{figure*}

\subsection{Individual detector transition probabilities}\label{subsec:indv-detector-prob-Ij}

In this subsection we study the individual detector transition probability $\mathcal{I}_{A_{\omega_{k}}}$, corresponding to detector $A$ in outgoing null path, through plots. We consider both the nonextremal and extremal scenarios, and the transition probability correspond to a certain field mode frequency $\omega_{k}$. For the explicit expression of $\mathcal{I}_{A_{\omega_{k}}}$ in the nonextremal and extremal scenarios one is referred to Eqs. (\ref{Appn:Ij-NE}) and (\ref{Appn:Ij-E}) of Appendix. The plots of $\mathcal{I}_{A_{\omega_{k}}}$ related to the nonextremal and extremal cases are given in the left side of Fig. \ref{fig:Ij-dE-NE-E}. From this figure it broadly seems that the extremal transition probability is lower than the nonextremal scenario. To further investigate this we have plotted the difference between the nonextremal and extremal transition probabilities in the right side of Fig. \ref{fig:Ij-dE-NE-E}. Then we observe that nonextremal detector transition is greater than extremal for all $\bar{\mathcal{E}}=\rs\,\Delta E$, when $\rq/\rs$ is low, e.g., when $\rq/\rs=0.1$ and $\rq/\rs=0.3$. It should be noted that when $\rq/\rs\to 1/2$ one observes the extremal scenario. On the other hand, near this extremal limit but still in the nonextremal case when $\rq/\rs= 0.4$, the transition is lower than the extremal case for moderately large $\bar{\mathcal{E}}\sim 0.3$. We also observe this feature becoming more prominent for higher $\rq/\rs$, such as for $\rq/\rs= 0.45$.

Therefore, depending on different black hole parameters like $\rq$, $\rs$, and detector transition energy $\Delta E$ one can observe more individual detector transition in the nonextremal or extremal cases of a \rn background. Near the extremal limit, from the nonextremal or extremal cases, each transition can be greater than the other depending on the value of $\bar{\mathcal{E}}$. While for low $\rq/\rs$ the nonextremal transition is greater than the extremal for all $\bar{\mathcal{E}}$. Thus one cannot predict a particular feature specific to the extremal and nonextremal scenarios for all parameter values, only in terms of the individual detector transition probabilities. This observation also motivates us to study the entanglement in these backgrounds and check whether the situation changes there.

\subsection{The measure of the entanglement: concurrence}\label{subsec:concurrence-NE-E}

\begin{figure*}
\centering
\includegraphics[width=0.475\linewidth]{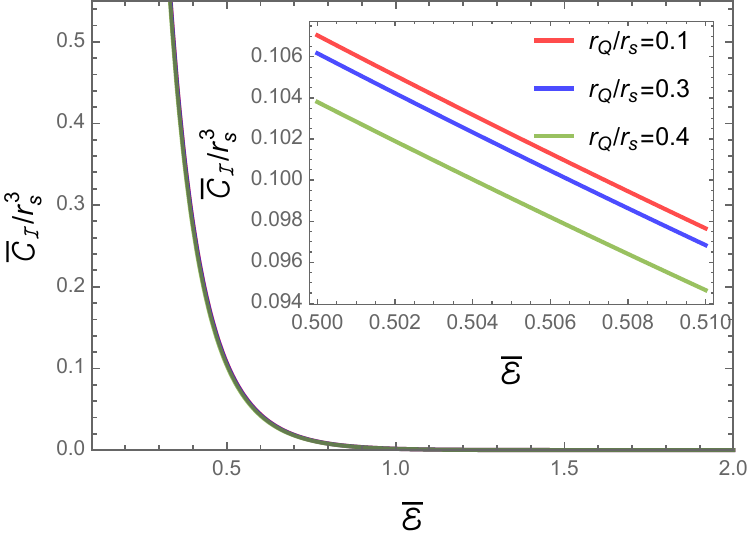}
\hskip 10pt
\includegraphics[width=0.47\linewidth]{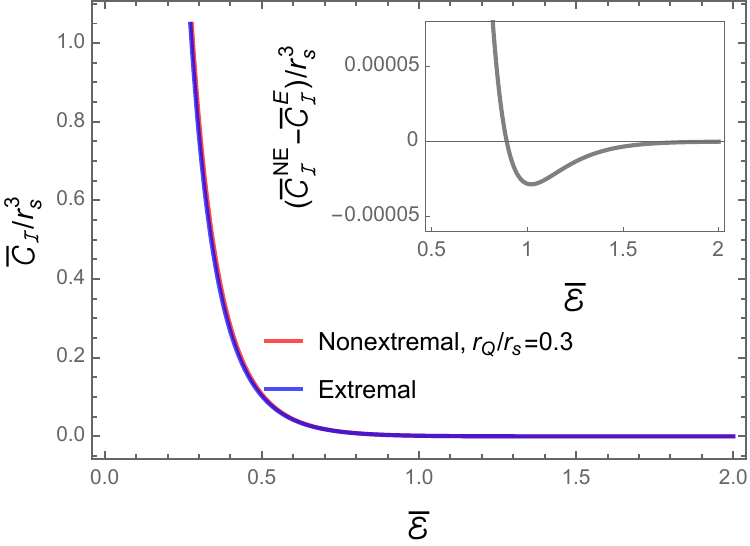}
\caption{The concurrence in the nonextremal and extremal cases are plotted as functions of the dimensionless energy gap $\bar{\mathcal{E}}=\rs\,\Delta E$. In the left figure the concurrence in the nonextremal scenario is depicted. While in the right figure the same for both nonextremal and extremal scenarios are illustrated. To obtain these plots we have fixed the other parameters at $\bar{d}_{n}=d_{n}/\rs=0$, $\bar{c}_{1}=c_{1}/\rs=0$, and $\bar{d}_{s}=d_{s}/\rs=1$. In right plot for the nonextremal case we have fixed $\bar{r}_{Q}=r_{Q}/\rs=0.3$. From these plots it is evident that in both the nonextremal and extremal cases, entanglement decreases with increasing $\bar{\mathcal{E}}$. In the epilogue of the right plot we have also depicted the difference of the concurrences between the nonextremal and extremal scenarios in gray.}\label{fig:conc-dE-NE-E}
\end{figure*}

\begin{figure}
\centering
\includegraphics[width=0.97\linewidth]{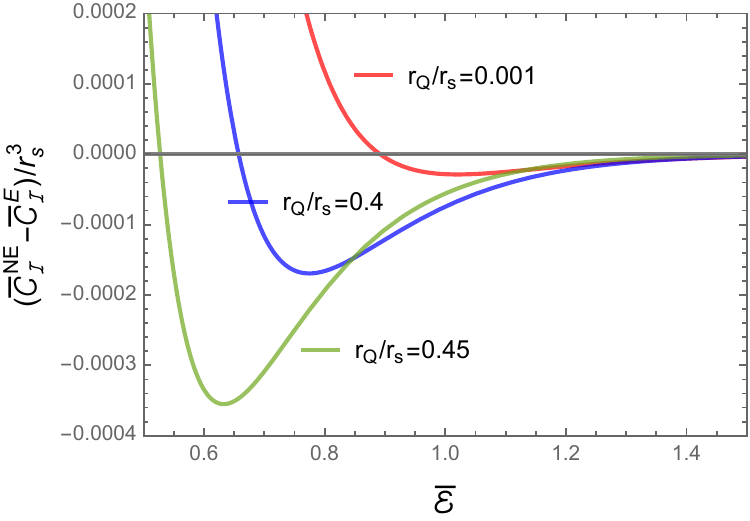}
\caption{The difference in concurrences between the nonextremal and extremal scenarios is plotted as a functions of $\bar{\mathcal{E}}$ for different  $\bar{r}_{Q}$. To obtain these plots we have fixed the other parameters at $\bar{d}_{n}=0$, $\bar{c}_{1}=0$, and $\bar{d}_{s}=1$. We find curves similar to that of $\bar{r}_{Q}=0.001$ for $\bar{r}_{Q}$ also many orders less than $0.001$. Therefore, in terms of concurrence, the difference between nonextremal and extremal scenarios broadly shows a specific feature for arbitrary $\bar{r}_{Q}$.}\label{fig:conc-dE-NE-E-diffs}
\end{figure}

Here we detail our findings on the measure of the entanglement, which is concurrence, in both the nonextremal and extremal scenarios. It is to be noted that for the static detector $B$ the integral $\mathcal{I}_{B_{\omega_{k}}}$ vanishes, which is evident from Eq. (\ref{eq:IB-RN-NE-1}). Then the multiplication of $\mathcal{I}_{A_{\omega_{k}}}$ and $\mathcal{I}_{B_{\omega_{k}}}$ will also vanish, for all finite $\mathcal{I}_{A_{\omega_{k}}}$. From Eqs. (\ref{Appn:Ij-NE}) and (\ref{Appn:Ij-E}) we observe for both the nonextremal and extremal cases that $\mathcal{I}_{A_{\omega_{k}}}$ are indeed finite. Then the concurrence is entirely given by $\mathcal{I}_{\varepsilon}$, see Eq. (\ref{eq:concurrence-I}). 

\begin{figure*}
\centering
\includegraphics[width=0.47\linewidth]{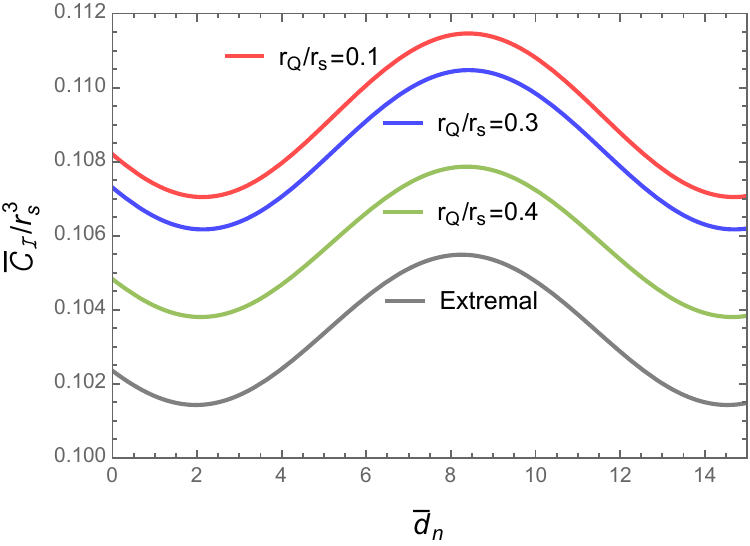}
\hskip 10pt
\includegraphics[width=0.48\linewidth]{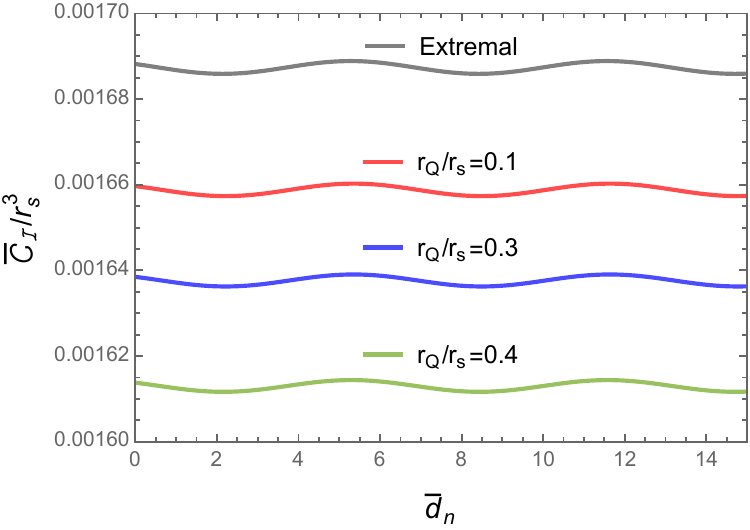}
\caption{The concurrence in the nonextremal and extremal cases are plotted as functions of $\bar{d}_{n}$. To obtain these plots we have fixed the other parameters at $\bar{c}_{1}=0$ and $\bar{d}_{s}=1$. The left plot corresponds to $\bar{\mathcal{E}}=0.5$ and the right plot corresponds to $\bar{\mathcal{E}}=1$.  With respect to the distance $\bar{d}_{n}$, that distinguishes different null paths, the concurrence in both cases are periodically dependant. In the left plot for $\bar{\mathcal{E}}=0.5$, the extremal curve has the lowest amplitude. While in the right plot for $\bar{\mathcal{E}}=1$, the extremal curve has the highest amplitude. This finding is consistent with our previous plots.}\label{fig:conc-ddn-NE-E}
\end{figure*}

\begin{figure*}
\centering
\includegraphics[width=0.47\linewidth]{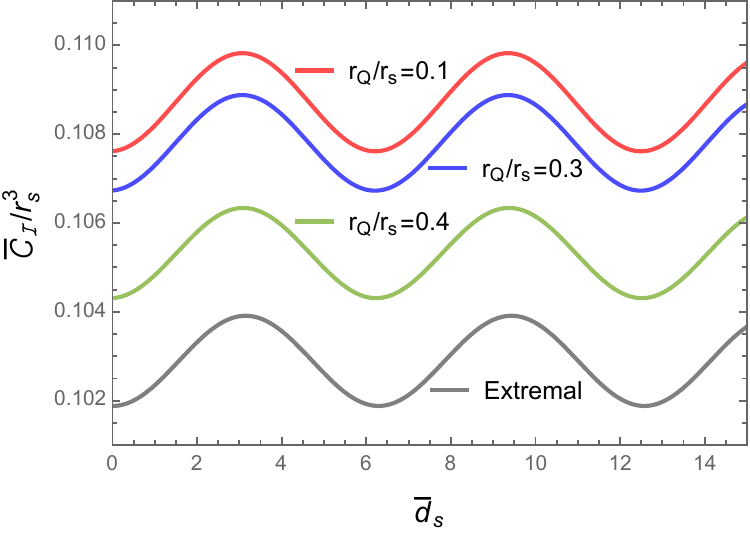}
\hskip 10pt
\includegraphics[width=0.475\linewidth]{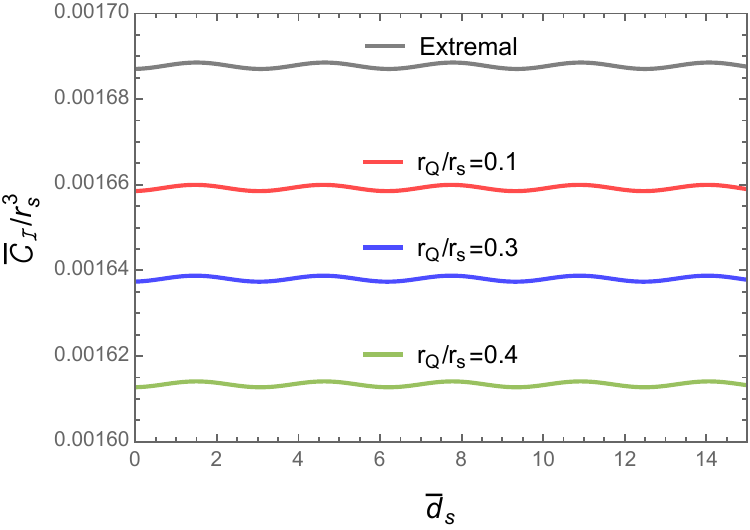}
\caption{The concurrence in the nonextremal and extremal cases are plotted as functions of $\bar{d}_{s}$. To obtain these plots we have fixed the other parameters at $\bar{d}_{n}=0$ and $\bar{c}_{1}=0$. In the left plot we have fixed $\bar{\mathcal{E}}=0.5$ and in the right plot $\bar{\mathcal{E}}=1$. With respect to the radial distance $\bar{d}_{s}$ of the static detector $B$, the concurrence in both the nonextremal and the extremal cases are periodic, and qualitatively the same.}\label{fig:conc-dds-NE-E}
\end{figure*}

\begin{figure}
\centering
\includegraphics[width=0.97\linewidth]{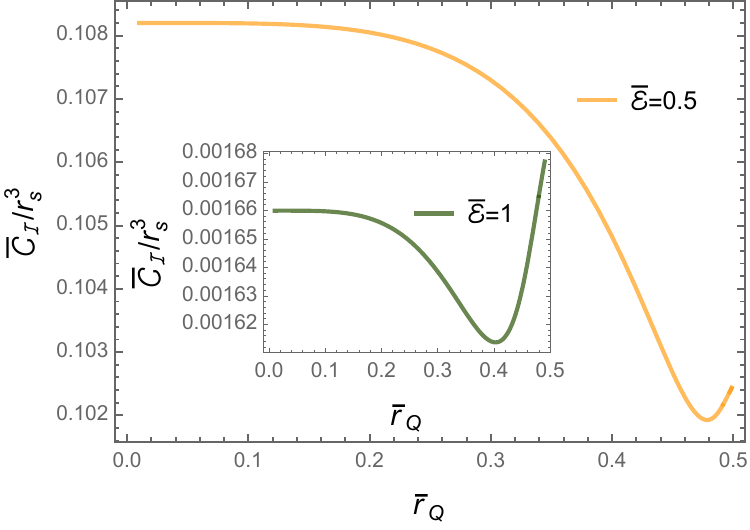}
\caption{The concurrence in the nonextremal case is plotted as a function of $\bar{r}_{Q}=\rq/\rs$. To obtain these plots we have fixed the other parameters at $\bar{d}_{n}=0$, $\bar{c}_{1}=0$, and $\bar{d}_{s}=1$. We fixed two different values of $\bar{\mathcal{E}}$ corresponding to two different curves.}\label{fig:conc-drQ-NE}
\end{figure}

It is now convenient to define the concurrence to be $\bar{\mathcal{C}}_{\mathcal{I}}=|\bar{\mathcal{I}}^{W}_{\varepsilon}(\Delta E)+\bar{\mathcal{I}}^{R}_{\varepsilon}(\Delta E)|$. Let us also define some dimensionless parameters of the system, with respect to which we shall study the characteristics of the concurrence. We define the dimensionless parameters $\bar{\mathcal{E}}=\rs\,\Delta E$, $\bar{r}_{Q}=r_{Q}/\rs$, $\bar{d}_{n}=d_{n}/\rs$, $\bar{c}_{1}=c_{1}/\rs$, and $\bar{d}_{s}=d_{s}/\rs$. Then the extremal limit is given by $\bar{r}_{Q}\to1/2$. With these considerations the dimensionless quantity that correspond to the concurrence is $\bar{\mathcal{C}}_{\mathcal{I}}/\rs^3$. In Fig. \ref{fig:conc-dE-NE-E}, \ref{fig:conc-dE-NE-E-diffs}, \ref{fig:conc-ddn-NE-E}, \ref{fig:conc-dds-NE-E}, and \ref{fig:conc-drQ-NE} we have plotted this quantity with respect to different system parameters.

\begin{figure}
\centering
\includegraphics[width=0.97\linewidth]{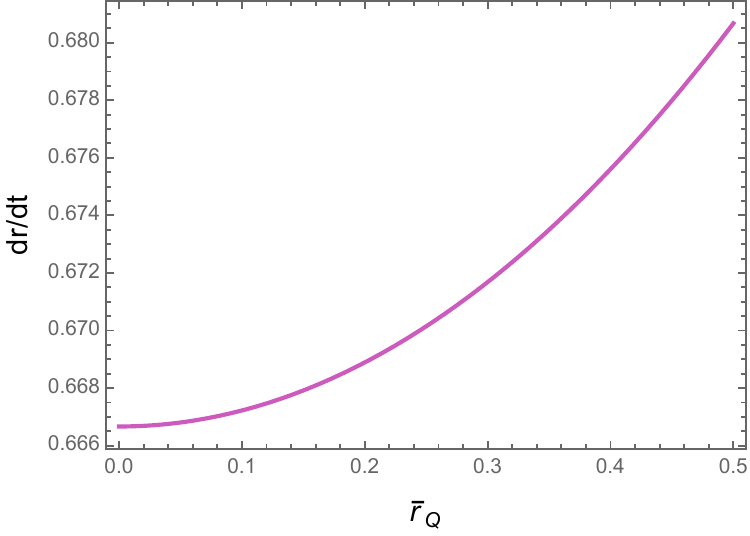}
\caption{$dr/dt$ from Eq. (\ref{eq:outgoing-null-path}) is plotted as a function of $\bar{r}_{Q}=\rq/\rs$. Here $\bar{r}_{Q}\to 1/2$ signifies the extremal limit $\rq \to\rs/2$, and in this limit the quantity $dr/dt$ tends to be the maximum. To obtain this plot we have fixed the radial position at $r/\rs = 5$. However, for any radial poistion outside the event horizon the characteristic of this curve remains the same.}\label{fig:dr-dt-vs-rQ}
\end{figure}

For instance, in Fig. \ref{fig:conc-dE-NE-E} we have plotted the concurrence as a function of the dimensionless detector energy gap $\bar{\mathcal{E}}$. In both the nonextremal (the left and right plots) and extremal (the right plot) cases  the entanglement decreases with increasing $\bar{\mathcal{E}}$, and eventually becomes vanishing at large $\bar{\mathcal{E}}$. However the nonextremal and extremal plots are not exactly the same. From the epilogue of the right plot of Fig. \ref{fig:conc-dE-NE-E} we observe that the nonextremal concurrence is larger than the extremal one at very low $\bar{\mathcal{E}}$. This gap decreases with increasing transition energy and becomes zero around $\bar{\mathcal{E}}\sim 0.8$, and then extremal concurrence becomes larger than the nonextremal one. Finally for very large $\bar{\mathcal{E}}\sim 1.5$ their difference is negligible and eventually the concurrences in both the cases become the same. Furthermore, in Fig. \ref{fig:conc-dE-NE-E-diffs} we have plotted the difference between the nonextremal and extremal concurrences as a function of the energy gap $\bar{\mathcal{E}}$ for different $\bar{r}_{Q}$. Here from very low $\bar{r}_{Q}$ ($\bar{r}_{Q}=0.001$) to near extremal cases ($\bar{r}_{Q}=0.45$) one observes that this difference is positive in low $\bar{\mathcal{E}}$. While the entanglement is larger in the extremal case for moderately larger $\bar{\mathcal{E}}$. Therefore, unlike the single detector case in entanglement one perceives a persistent feature in the nonextremal and extremal cases for different black hole charges. This feature dictates that with very low detector transition energy one can get more entanglement from the nonextremal background. Whereas, with moderately large transition energy one is able to get maximum entanglement from the extremal background. It should be mentioned that, similar to the concurrence, in the single detector transition these features were not maintained thoroughly for arbitrary black hole charges, see the right plots of Fig. \ref{fig:Ij-dE-NE-E} and the discussion of \ref{subsec:indv-detector-prob-Ij}.

On the other hand, in Fig. \ref{fig:conc-ddn-NE-E} we have plotted the concurrence as a function of the dimensionless distance $\bar{d}_{n}$, that distinguishes different outgoing null paths. We observe that the entanglement is periodic with respect to this distance in both the nonextremal and the extremal scenarios. The periodicity and amplitude depend on the value of the detector transition energy $\bar{\mathcal{E}}$, as is perceived from the left and the right plots. At higher transition energy the amplitude is lower and the periodicity is greater. We also observe that for low detector transition energy $\bar{\mathcal{E}}=0.5$ (the left plot) the extremal case provides entanglement lower than the nonextremal case. While for high transition energy $\bar{\mathcal{E}}=1$ (the right plot) the extremal case provides more entanglement than the nonextremal one. This particular finding is also consistent with the prediction from the epilogue of the right plot of Fig. \ref{fig:conc-dE-NE-E}.

From Eqs. (\ref{eq:IeW-RN-NE-4}, \ref{eq:IeR-RN-NE-6}) of the nonextremal case and Eqs. (\ref{eq:IeW-RN-E-1}, \ref{eq:IeR-RN-E-1}) of the extremal case one can observe that there is an overall phase factor containing the parameter $\bar{c}_{1}$ in the integrals defining the concurrence. Now as the concurrence $\bar{\mathcal{C}}_{\mathcal{I}}$ is obtained from the modulus of these integrals, it will not depend on $\bar{c}_{1}$ . Therefore, in both the nonextremal and extremal cases the concurrence is independent of $\bar{c}_{1}$.

In Fig. (\ref{fig:conc-dds-NE-E}) we have plotted the concurrence as a function of the dimensionless distance $\bar{d}_{s}$ of the static detector. For both the nonextremal and the extremal cases, the concurrence is found to be periodically dependent on $\bar{d}_{s}$, and qualitatively are the same. Here also the periodicity and amplitude of the concurrence depend on $\bar{\mathcal{E}}$, see the left and the right plots. For higher detector transition energy the periodicity is greater and the amplitude is lower. For low detector transition energy $\bar{\mathcal{E}}=0.5$ (the left plot) the extremal entanglement is lower than the nonextremal case. While for high transition energy $\bar{\mathcal{E}}=1$ (the right plot) the extremal entanglement is greater than the nonextremal one.

In Fig. \ref{fig:conc-drQ-NE} we plotted the concurrence in the nonextremal scenario as a function of the dimensionless charge $\bar{r}_{Q}$ of the black hole. The concurrence first decreases, and then increases with increasing $\bar{r}_{Q}$, as the charge $\bar{r}_{Q}\to 1/2$, i.e., the extremal limit. For $\bar{\mathcal{E}}=0.5$, the concurrence does not become greater than all the nonextremal entanglement in the extremal limit. On the other hand, for $\bar{\mathcal{E}}=1$, the concurrence in the extremal limit is seem to be always greater than the nonextremal case. This observation further solidifies our perception from the previous plots, that for large detector transition energy one may get more entanglement from the extremal background than the nonextremal case.

With these above observations one can obtain a perception on the quantitative distinction between the nonextremal and extremal Reissner-Nordstr\"om black hole backgrounds in terms of the entanglement with one static Unruh-DeWitt detector and another one in outgoing null trajectory.

\vspace{0.2cm}
\section{Discussion}\label{sec:discussion}
The concept of extremality is quite intriguing in a black hole spacetime. An extremal black hole does not Hawking radiate, though possessing an event horizon. Thus extremal black holes are qualitatively different from nonextremal ones. It is also suggested in the literature \cite{Ghosh:1996gp, Gao:2002kz, Pradhan:2012yx} that to talk about the characteristics of an extremal black hole, one should start with an extremal one rather than taking the extremal limit of the nonextremal one. In this work, we have studied the characteristics of the entanglement profiles in nonextremal and extremal \rn black hole spacetimes. In this regard, we have considered detector $A$ in outgoing null trajectory and a static detector $B$. Our first observation is that the individual detector transition probability of detector $A$ corresponding to the Boulware like vacuum is non-vanishing in both nonextremal and extremal cases, see Fig. \ref{fig:Ij-dE-NE-E}. We believe that the motion of the detector stimulates this nonzero transition probability also in the extremal case. In terms of this detector transition probability there is no specific feature in the difference between the nonextremal and the extremal scenarios for arbitrary black hole charge $\rq$, see the right plots of Fig \ref{fig:Ij-dE-NE-E} and the discussion in Sec. \ref{subsec:indv-detector-prob-Ij}. Furthermore, it will be interesting to construct suitable Kruskal-like coordinates in an extremal \rn black hole spacetime \cite{Gao:2002kz} and check the response of the static detector corresponding to the Unruh or Hartle-Hawking like vacuum. One should expect this detector response to be vanishing according to Hawking quanta's vanishing number density in the extremal scenario. We are presently working in this direction and intend to present our findings in a future communication. We also observed that the individual detector transition probability (rather the rate of this transition probability) vanishes for the static detector $B$, corresponding to the Boulware like vacuum.
\vspace{0.2cm}


We now discuss and understand the outcomes obtained in entanglement profiles in the extremal and nonextremal scenarios. Our observations are as follows.

\begin{itemize}
\item First, we observe that the extremal and nonextremal scenarios provide qualitatively the same results regarding entanglement profiles. We have seen that one can obtain the extremal tortoise coordinate by carefully taking the extremal limit from the nonextremal case. However, how to handle this same limit in the nonextremal nonlocal term of entanglement is yet to be discovered (see Sec. \ref{sec:RN-BH-spacetime} and Appendix \ref{Appn:tortoise-NE-to-E}). Therefore, even if one evaluates the concurrence numerically in the nonextremal scenario, the extremal limit from that result may not work out. The most plausible thing to do is to consider the extremal and the nonextremal cases separately from the beginning for entanglement. Our analysis also considers the extremal and the nonextremal cases separately from the beginning. In this work, we provided a rigorous mathematical understanding of the nonextremal and extremal entanglement profiles. Furthermore, even in terms of entanglement profiles, we established that the extremal scenario, as expected, is quantitatively a limiting case for the nonextremal one.

\item Second, we observe that the entanglement decreases with increasing detector transition energy for both the nonextremal and extremal black holes. In this regard, one can refer to Fig. \ref{fig:conc-dE-NE-E}. This result is quite natural as physically detectors with lower transition energy should need lesser energy from the background field to get excited and to get entangled.

\item Third, unlike the individual detector transition (say for detector $A$), the quantitative difference in concurrence between the nonextremal and extremal scenarios exhibits specific features for a broad range of black hole charges $\rq$.

\begin{itemize}
\item[A.] For instance, this difference suggests one can get more entanglement from the nonextremal background than the extremal one when the detector transition energy is very low. While for moderately high detector transition energy extremal case showcases maximum entanglement (Figs. \ref{fig:conc-dE-NE-E}, \ref{fig:conc-dE-NE-E-diffs}, \ref{fig:conc-drQ-NE}).

\item[B.] To understand the physical reasoning behind this phenomenon of greater entanglement from a nonextremal background in a low transition energy regime, let us look into the plot of $dr/dt$ (Eq. (\ref{eq:outgoing-null-path})) vs. $\rq$ in Fig. \ref{fig:dr-dt-vs-rQ}. From this plot, we observe that as $\rq$ increases and tends to reach $\rs/2$ (the extremal limit), the quantity $dr/dt$, which corresponds to the momentum of an observer in an outgoing null path, increases. This should have an effect on the entanglement profile also. It may happen since, in the extremal scenario, the momentum of the detector is higher than in the nonextremal case. Thus the energy associated with the detectors in the extremal case is higher. Therefore, in this case, entanglement is possible at higher energy.

\end{itemize}

\item Fourth, in Fig. \ref{fig:conc-ddn-NE-E}, we observed that both the nonextremal and extremal entanglement are periodic with respect to the distance $d_{n}$ of the null paths of the detector $A$. Furthermore, from Fig. \ref{fig:conc-dds-NE-E}, a similar periodic nature is observed with respect to the distance $d_{s}$ of the static detector $B$. A similar periodic nature in the entanglement profile was also observed in \cite{Barman:2021kwg}. However, unlike \cite{Barman:2021kwg}, where both of the detectors were in outgoing null paths, we do not observe any entanglement shadow regions or points here. Our observations suggest these periodicities are dependent on the detector transition energy. For instance, the periodicity increases with increasing transition energy. At the same time, the amplitude of this oscillation of concurrence decreases with increasing detector transition energy.

\item Therefore, with our considered set-up and in terms of entanglement, the nonextremal and extremal cases are qualitatively indistinguishable. However, quantitatively they differ. Our observations suggest that for arbitrary fixed detector transition energies, the optimum entanglement is achieved depending on the extremal or nonextremal scenarios. Let us elucidate this finding with a hypothetical example. In this regard, let us consider there are two RN backgrounds, one of them being extremal and another nonextremal. However, an outside observer does not know which is extremal or nonextremal. Then our observations state that by investigating the entanglement profiles in these two backgrounds for varying detector transition energies, one can figure out which is extremal or nonextremal. As one increases the detector transition energy, the extremal background will start to report more entanglement than the nonextremal one. We should also mention that to carry out this experiment one must also have the knowledge about the black hole masses. Otherwise the scaling of the concurrences in terms of $r_{s}^3$ would not be possible, see Fig. \ref{fig:conc-drQ-NE}. A different condition that does not require the prior knowledge of the black hole masses is if both the black holes have the same mass. In that scenario both nonextremal and extremal concurrences are scaled by the same quantity, and thus are comparable.
 
\end{itemize}

We would want to mention that if Alice followed a timelike trajectory, we expect there would have been qualitative changes in the entanglement, as entanglement features depend on specific detector trajectories. Furthermore, we also expect that with Alice following timelike trajectory some of our key observations may not change. Such observations include: (I) the extremal entanglement being a limit of the nonextremal one, (II) with varying detector energy gap the extremal entanglement can become maximum compared to nonextremal scenarios. 
To elaborate this, let us consider a radial outgoing observer with velocity $u^{a}=\left(dt_{c}/d\bar{\lambda},\,dr/d\bar{\lambda}\right)=(1/\mathcal{F}(r),\sqrt{1-\mathcal{F}(r)})$, with $\bar{\lambda}$ being a suitable affine parameter and $\mathcal{F}(r)=1-(r_{s}/r)+(r_{Q}^2/r^2)$. One can check that this denotes a timelike observer as $g_{ab}u^{a}u^{b}=-1$. From the expressions of $dr/d\bar{\lambda}$ and $dt_{c}/d\bar{\lambda}$ one can obtain the detector trajectory, which can then be utilized to obtain the entanglement. However, we believe the analytical evaluation of entanglement with this timelike trajectory would not be possible, and one needs to take a numerical approach. At the same time, one can gain some physical insight about the nature of entanglement from the expression of $dr/dt_{c}=\mathcal{F}(r)\sqrt{1-\mathcal{F}(r)}$. In particular, for a fixed radius $r$ this quantity keeps increasing as $r_{Q}$ approaches $r_{s}/2$. Therefore, one may expect that this can have some physical implications in the entanglement measure as the above quantity relates to detector energy in that particular trajectory. However, to know the exact outcome one needs to thoroughly investigate this situation.

Finally, we would like to mention (we observed it, though not included in the manuscript) that two static detectors do not provide any entanglement from the Boulware-like vacuum of the \rn black hole background. 
%
This is analogous to the phenomenon of static detectors not providing entanglement from the Minkowski vacuum for eternal switching \cite{Koga:2018the}. However, if a finite switching is considered the static detectors in our scenario may also start giving entanglement like the detectors in flat background.
%
%
Moreover, one may also consider both of the detectors to be in null trajectories like done in \cite{Barman:2021kwg}. But in that case, evaluating the nonlocal entangling term becomes a challenge analytically. We would like to pursue these issues in the future.\vspace{0.6cm}


\vspace{0.2cm}
\begin{acknowledgments}
SB would like to thank the Science and Engineering Research Board (SERB), Government of India (GoI), for supporting this work through the National Post Doctoral Fellowship (N-PDF, File number: PDF/2022/000428). The research of BRM is partially supported by a START-UP RESEARCH GRANT (No. SG/PHY/P/BRM/01) from the Indian Institute of Technology Guwahati, India.
\end{acknowledgments}
\vspace{0.2cm}


\appendix
\begin{appendix}
\begin{widetext}

\section{Nonextremal tortoise coordinate in the limit $\rq\to\rs/2$}\label{Appn:tortoise-NE-to-E}

In order to realize the limit $\rq\to\rs/2$ of the tortoise coordinate in Eq. (\ref{eq:tortoise-NE}) of the nonextremal case and thus to get the extremal expression (\ref{eq:tortoise-E}), we consider $\rs^2-4\rq^2 = \rs^2\,\varsigma^2$. Then $\varsigma \to 0$ should represent the extremal scenario. With this consideration one has $r_{\pm} = (1\pm\varsigma )\,\rs/2$ and $\kappa_{\pm} = 2 \varsigma/[\rs(1\pm\varsigma)^2]$. Then around $\varsigma=0$ one can approximate the quantity 
\begin{eqnarray}\label{eq:tortoise-NE2}
\ln \Big(\frac{r}{r_{\pm}}-1\Big) &=& \ln \Big(\frac{2r}{\rs(1\pm\varsigma)}-1\Big)\nonumber\\
~&\approx& \ln \Big(\frac{2r}{\rs}(1\mp\varsigma)-1\Big)\nonumber\\
~&=& \ln \Big(\frac{2r}{\rs}-1\Big) + \ln \Big(1\mp\frac{2r\varsigma}{2r-\rs}\Big) \nonumber\\
~&\approx& \ln \Big(\frac{2r}{\rs}-1\Big) \mp\frac{2r\varsigma}{2r-\rs}~.
\end{eqnarray}
It should be noted that in the last expression we have used the approximation $\ln (1\mp\delta) \approx \mp\delta$, for very small positive $\delta$. Now one may use the previous expression in (\ref{eq:tortoise-NE}) to get 
\begin{eqnarray}\label{eq:tortoise-NE3}
\rstar &\approx& r+\frac{\rs\varsigma}{\varsigma} \ln \Big(\frac{2r}{\rs}-1\Big)-\frac{r\rs(1+\varsigma^2)}{2r-\rs}\nonumber\\
~&=& r+\rs \ln \Big(\frac{2r}{\rs}-1\Big)-\frac{(\rs^2/4)(1+\varsigma^2)}{r-\rs/2}-\frac{\rs(1+\varsigma^2)}{2}~.\nonumber\\
\end{eqnarray}
In the limit of $\varsigma\to 0$ this expression looks same as the extremal expression of the tortoise coordinate (\ref{eq:tortoise-E}) up-to a constant additive quantity $(-\rs/2)$. One should notice that due to the presence of both the $(1/2\kappa_{\pm})\,\ln (r/r_{\pm}-1)$ terms in the expression of the nonextremal tortoise coordinate (\ref{eq:tortoise-NE}), there was this appearance of zero divided by zero in the extremal limit. This zero by zero form is evident from the first expression of Eq. (\ref{eq:tortoise-NE3}), which does not create too much hurdle to get the extremal expression (\ref{eq:tortoise-E}) eventually. However, one should note that during further calculations of $\mathcal{I}_{j}$ and $\mathcal{I}_{\varepsilon}$ when the terms $(1/2\kappa_{\pm})\,\ln (r/r_{\pm}-1)$ are not kept together the limit to extremality may not occur as easily. Therefore, it seems necessary to consider the nonextremal and extremal scenarios separately from the beginning to estimate the integrals $\mathcal{I}_{j}$ and $\mathcal{I}_{\varepsilon}$.

\section{Affine parameter for a detector in outgoing null trajectory}\label{Appn:Affine-null}

Here we elucidate how the Eddington-Finkelstein (EF) time coordinate can be interpreted as one of the affine parameters corresponding to an observer in outgoing null trajectory. To understand this we first recall the expression of the line element in EF coordinates from Eq. \eqref{eq:metric-RN-EF}
\begin{eqnarray}\label{eq:EF-metric}
ds^2 = -\mathcal{F}(r)\,dt^2+2\,\{1-\mathcal{F}(r)\}\,dtdr+\{2-\mathcal{F}(r)\}\,dr^2~,
\end{eqnarray}
where, $\mathcal{F}(r)=1-(r_{s}/r)+(r_{Q}^2/r^2)$. We also have the EF time as $t=t_{c}+r_{\star}-r$, with $t_{c}$ being the Reissner-Nordstr\"om coordinate time, $r$ the radial coordinate, and $r_{\star}$ denoting the tortoise coordinate defined as $dr_{\star}=dr/\mathcal{F}(r)$. One can obtain the null trajectories by putting $ds^2=0$ in Eq. \eqref{eq:EF-metric}. In particular, for a an outgoing null path the relation between the EF time and the radial coordinate is obtained from
\begin{eqnarray}\label{eq:EF-time-null}
\frac{dt}{dr} = \frac{2-\mathcal{F}(r)}{\mathcal{F}(r)}~.
\end{eqnarray}

In the spacetime described by the EF coordinates from Eq. \eqref{eq:EF-metric}, we consider a massless particle following a null trajectory with momenta $P_{a}$, where $a$ denotes the time and space indices. Then it is straightforward to understand that $E=-P_a\chi^a=-P_{t}$ signifies the energy of the particle as the timelike Killing vector $\chi^a$ is given by $\chi^a=(1,0)$. With the help of the above metric the dispersion relation for this massless particle, $g^{ab}P_{a}P_{b}=0$, becomes
\begin{eqnarray}\label{eq:dispersion}
-\{2-\mathcal{F}(r)\}\,E^2-2\{1-\mathcal{F}(r)\}EP_{r}+\mathcal{F}(r)P_{r}^2=0~.
\end{eqnarray}
We should also mention that the contravariant components of the momentum $P^{a}=dx^{a}/d\lambda$ signifies the velocity with $\lambda$ being a suitable affine parameter, which is yet to be determined. From the above equation we obtain
\begin{eqnarray}\label{eq:energy}
E=\frac{P_{r}}{2-\mathcal{F}(r)}\,(\mathcal{F}(r)-1\pm 1)~.
\end{eqnarray}
In the above expression the $``\pm"$ sign correspond to the outgoing and ingoing null trajectories. In particular, for the outgoing trajectory the equation of motion results in
\begin{eqnarray}\label{eq:affime-para}
\frac{dr}{d\lambda} = \frac{\partial E}{\partial P_{r}}=\frac{\mathcal{F}(r)}{2-\mathcal{F}(r)}~,
\end{eqnarray}
which can be integrated to obtain the affine parameter in this trajectory. Now one can notice the striking resemblance between the EF time from Eq. \eqref{eq:EF-time-null} and the affine parameter from Eq. \eqref{eq:affime-para} both of which correspond to an observer in outgoing null path. In particular, in the second scenario, the the affine parameter was arbitrary, but turns out to be the same as the EF time $t$ as obtained from Eq. \eqref{eq:EF-time-null} corresponding to an observer in null path.

\section{Evaluation of $\mathcal{I}_{j}$ and $\mathcal{I}_{\varepsilon}$ in nonextremal \rn spacetime}\label{Appn:Ij-Ie-NE}

\subsection{Evaluation of $\mathcal{I}_{A_{\omega_{k}}}$}\label{Appn:Ij-NE}

In Eq. (\ref{eq:IA-RN-NE-3}) let us consider that $\mathcal{I}_{A_{\omega_{k}}} = (\rp^2/4\pi\omega_{k})\,|\bar{\mathcal{I}}_{A_{\omega_{k}}}|^2$. We shall now provide some formulas and the decomposition of the integrand inside $\bar{\mathcal{I}}_{A_{\omega_{k}}}$, using which one can reach an expression of $\bar{\mathcal{I}}_{A_{\omega_{k}}}$ in a methodical and simple manner. In this regard, let us specify the integral representation 
%
\begin{eqnarray}\label{eq:AppnB-NE-2}
\int_{0}^{\infty}dx\,x^{a+n} e^{-z\, x} (\delta +x)^{-a+b-1} &=& \pi  \csc (\pi  (b+n)) \Bigg[z^{-b-n} \, _1\tilde{F}_1(a-b+1;-b-n+1;z \delta )\nonumber\\
~&& -\frac{\Gamma (a+n+1) \delta ^{b+n} \, _1\tilde{F}_1(a+n+1;b+n+1;z \delta )}{\Gamma (a-b+1)}\Bigg]\,,
\end{eqnarray}
%
which is obtained only when the conditions $Real[a]>0$, $Real[z]>0$, $Real[\delta]>0$, and $n\ge -1$ are satisfied. For reference, we should mention that similar integral representations can also be found in \cite{gradshteyn2007}, page $348$ and $1023$. In this expression the function $ _1\tilde{F}_1(a;b;z)$ denotes the regularized confluent hypergeometric function, which is related to the confluent hypergeometric function via the relation $ _1\tilde{F}_1(a;b;z)\,=\, _1 F_1(a;b;z)/\Gamma(b)$ with $\Gamma(b)$ denoting the Gamma function. Now let us check whether the integral $\bar{\mathcal{I}}_{A_{\omega_{k}}}$ from Eq. (\ref{eq:IA-RN-NE-3}) can be cast into a form similar to the representation of Eq. (\ref{eq:AppnB-NE-2}). In particular, this integral $\bar{\mathcal{I}}_{A_{\omega_{k}}}$ looks like
\begin{eqnarray}\label{eq:AppnB-NE-2p}
\bar{\mathcal{I}}_{A_{\omega_{k}}} = \int_{0}^{\infty} dy_{A}\, 
\frac{2(y_{A}+1)^2\rp}{y_{A}\big[(y_{A}+1)\rp-\rm\big]} \,\frac{e^{i\rp(\Delta E+2\omega_{k})y_{A}}}{y_{A}^{-\frac{i(\Delta 
E+\omega_{k})}{\kp}}}\, \bigg[\frac{(y_{A}+1)\rp}{\rm}-1\bigg]^{-\frac{i(\Delta 
E+\omega_{k})}{\km}}~.
\end{eqnarray}
After introducing the regulator $e^{-\epsilon \,y_{A}}y_{A}^{\epsilon}$, we consider a change of variables $x=y_{A}\rp/\rm$, $a=\epsilon+i(\Delta E+\omega_{k})/\kp$, $b=a-i(\Delta E+\omega_{k})/\km$, $z=\{\epsilon-i(\Delta E+2\omega_{k})\rp\}\rm/\rp$, and $\delta= \rp/\rm-1$. Then the integral $\bar{\mathcal{I}}_{A_{\omega_{k}}}$ from Eq. (\ref{eq:AppnB-NE-2p}) can be represented as 
\begin{eqnarray}\label{eq:AppnB-NE-3}
\bar{\mathcal{I}}_{A_{\omega_{k}}} = 2\left(\frac{\rm}{\rp}\right)^{-1+a} \int_{0}^{\infty}dx\,\left(\frac{x^2\rm^2}{\rp^2}+\frac{2x\rm}{\rp}+1\right)\,x^{a-1} e^{-z\, x} (\delta +x)^{-a+b-1}~,
\end{eqnarray}
where all the redefined parameters $a$, $z$, and $\delta$ satisfy the necessary conditions to utilize the integral representation from Eq. (\ref{eq:AppnB-NE-2}). 
Here the relevance of the considered regulator with a positive real $\epsilon$ becomes apparent. Therefore, using the expression from (\ref{eq:AppnB-NE-2}) one can obtain the result for the integral $\bar{\mathcal{I}}_{A_{\omega_{k}}}$ in a straightforward manner. In particular, in terms of the redefined parameters this expression looks like 
\begin{eqnarray}\label{eq:AppnB-NE-1}
\bar{\mathcal{I}}_{A_{\omega_{k}}} &=& 2\left(\frac{\rm}{\rp}\right)^{-1+a} \times \frac{\pi  \csc (\pi  b)}{\delta  \rp^2 z} \Bigg[\frac{z^{-b}}{a} \bigg\{\rp z \, _1\tilde{F}_1(a-b+1;1-b;z \delta ) (2 a \delta  \rm-b \rp+\delta  \rp z) \nonumber\\
~&& -\, _1\tilde{F}_1(a-b+1;-b;z \delta ) \left(a \delta  \rm^2+\rp^2 z\right)\bigg\}\nonumber\\
~&& +\frac{\Gamma (a) \delta ^{b}}{\Gamma (a-b+1)} \bigg\{\, _1\tilde{F}_1(a;b;z \delta ) \left(\delta  \rm^2 (a-b)+z (\rp-\delta  \rm)^2\right)\nonumber\\
~&& +\delta  \rm (a-b) \, _1\tilde{F}_1(a;b+1;z \delta ) (-b \rm+\delta  \rm z-2 \rp z)\bigg\}\Bigg]~.
\end{eqnarray}

\subsection{Evaluation of $\mathcal{I}_{\varepsilon_{\omega_{k}}}$}\label{Appn:Ie-NE}

In order to evaluate the quantity $\bar{\mathcal{I}}^{W}_{\varepsilon}(\omega_{k})$ from Eq. (\ref{eq:IeW-RN-NE-4}) corresponding to the nonextremal RN scenario we first introduce regulator $e^{-\epsilon \,y_{A}}y_{A}^{\epsilon}$ and consider a change of variables $x=y_{A}\rp/\rm$, $a=\epsilon+i(\Delta E+\omega_{k})/\kp$, $b=a-i(\Delta E+\omega_{k})/\km$, $z=\{\epsilon-i(\Delta E+2\omega_{k})\rp\}\rm/\rp$, and $\delta= \rp/\rm-1$, as previously mentioned. Then the integral $\bar{\mathcal{I}}^{W}_{\varepsilon}(\omega_{k})$ from Eq. (\ref{eq:IeW-RN-NE-4}) becomes
\begin{eqnarray}\label{eq:AppnB-NE-4}
\bar{\mathcal{I}}^{W}_{\varepsilon}(\omega_{k}) &=& \frac{\rp\rs}{\omega_{k}}\,e^{i\omega_{k} \,(d_{n}-c_{1}-d_{s})}\,e^{i\Delta E \,d_{n}}\,e^{i\rp(\Delta E+2\omega_{k})}\nonumber\\
~&\times&~\left(\frac{\rm}{\rp}\right)^{-1+a} \int_{0}^{\infty}dx\,\left(\frac{x^2\rm^2}{\rp^2}+\frac{2x\rm}{\rp}+1\right)\,x^{a-1} e^{-z\, x} (\delta +x)^{-a+b-1}\nonumber\\
~&=& \frac{\rp\rs}{2\omega_{k}}\,e^{i\omega_{k} \,(d_{n}-c_{1}-d_{s})}\,e^{i\Delta E \,d_{n}}\,e^{i\rp(\Delta E+2\omega_{k})}~\bar{\mathcal{I}}_{A_{\omega_{k}}}~.
\end{eqnarray}
Therefore, using the expression of $\bar{\mathcal{I}}_{A_{\omega_{k}}}$ from Eq. (\ref{eq:AppnB-NE-1}) one can easily obtain the expression of $\bar{\mathcal{I}}^{W}_{\varepsilon}(\omega_{k})$ utilizing the previous relation (\ref{eq:AppnB-NE-4}).

Next to evaluate $\bar{\mathcal{I}}^{R}_{\varepsilon}(\omega_{k})$ from Eq. (\ref{eq:IeR-RN-NE-6}) corresponding to the nonextremal case, let us express it as $\bar{\mathcal{I}}^{R}_{\varepsilon}(\omega_{k})=\bar{\mathcal{I}}^{R^{+}}_{\varepsilon}(\omega_{k})+\bar{\mathcal{I}}^{R^{-}}_{\varepsilon}(\omega_{k})$. From (\ref{eq:IeR-RN-NE-6}) one can obtain the expressions of  these $\bar{\mathcal{I}}^{R^{\pm}}_{\varepsilon}(\omega_{k})$ as
\begin{eqnarray}\label{eq:AppnB-NE-IeR-1}
\bar{\mathcal{I}}^{R^{\pm}}_{\varepsilon}(\omega_{k}) &=& -\frac{\rp^2\rs\,e^{i\Delta E(d_{n}-c_{1})}}{2\,\omega_{k}\rm} \,e^{2i\rp\Delta E\pm i\omega_{k}d_{s}\mp i\rp\omega_{k}}\,\int_{0}^{\infty} dy_{A}\, 
(y_{A}+1)^2\nonumber\\
~&\times& y_{A}^{-1+\frac{i}{2\kp}(3\Delta 
E\mp \omega_{k})}\,e^{i\rp(2\Delta E\mp \omega_{k})y_{A}}\, \bigg[\tfrac{(y_{A}+1)\rp}{\rm}-1\bigg]^{-1-\frac{i}{2\km}(3\Delta E\mp \omega_{k})}~.
\end{eqnarray}
Now we introduce regulator of the form $e^{-\epsilon \,y_{A}}y_{A}^{\epsilon}$, and consider a change of variables similar to the previous cases as $x=y_{A}\rp/\rm$, $a'_{\pm}=\epsilon+i(3\Delta E\mp\omega_{k})/(2\kp)$, $b'_{\pm}=a'_{\pm}-i(3\Delta E\mp\omega_{k})/(2\km)$, $z'_{\pm}=\{\epsilon-i(2\Delta E\mp\omega_{k})\rp\}\rm/\rp$, and $\delta= \rp/\rm-1$. Here also one can perceive that the conditions $Real[a'_{\pm}]>0$, $Real[z'_{\pm}]>0$, and $Real[\delta]>0$ are satisfied. Moreover, one can express the previous integrals as
\begin{eqnarray}\label{eq:AppnB-NE-IeR-2}
\bar{\mathcal{I}}^{R^{\pm}}_{\varepsilon}(\omega_{k}) &=& -\frac{\rp\rs\,e^{i\Delta E(d_{n}-c_{1})}}{2\,\omega_{k}} \,e^{2i\rp\Delta E\pm i\omega_{k}d_{s}\mp i\rp\omega_{k}}\,\left(\frac{\rm}{\rp}\right)^{-1+a'_{\pm}}\nonumber\\
~&&\times~ \int_{0}^{\infty}dx\,\left(\frac{x^2\rm^2}{\rp^2}+\frac{2x\rm}{\rp}+1\right)\,x^{a'_{\pm}-1} e^{-z'_{\pm}\, x}~ (\delta +x)^{-a'_{\pm}+b'_{\pm}-1}~,
\end{eqnarray}
which can be evaluated using the integral representation from Eq. (\ref{eq:AppnB-NE-2}). Nevertheless let us explicitly express the forms of these $\bar{\mathcal{I}}^{R^{\pm}}_{\varepsilon}(\omega_{k})$ in terms of the above redefined parameters. For instance, the explicit expressions of $\bar{\mathcal{I}}^{R^{\pm}}_{\varepsilon}(\omega_{k})$ are given by 
\begin{eqnarray}\label{eq:AppnB-NE-IeR-3}
\bar{\mathcal{I}}^{R^{\pm}}_{\varepsilon}(\omega_{k}) &=& -\frac{\rp\rs\,e^{i\Delta E(d_{n}-c_{1})}}{2\,\omega_{k}} \,e^{2i\rp\Delta E\pm i\omega_{k}d_{s}\mp i\rp\omega_{k}}\,\left(\frac{\rm}{\rp}\right)^{-1+a'_{\pm}}\nonumber\\
~&\times&~ \frac{\pi  \csc (\pi  b'_{\pm})}{\delta  \rp^2 z'_{\pm}} \Bigg[\frac{z_{\pm}^{\prime-b'_{\pm}}}{a'_{\pm}} \bigg\{\rp z'_{\pm} \, _1\tilde{F}_1(a'_{\pm}-b'_{\pm}+1;1-b'_{\pm};z'_{\pm} \delta ) (2 a'_{\pm} \delta  \rm-b'_{\pm} \rp+\delta  \rp z'_{\pm}) \nonumber\\
~&& -\, _1\tilde{F}_1(a'_{\pm}-b'_{\pm}+1;-b'_{\pm};z'_{\pm} \delta ) \left(a'_{\pm} \delta  \rm^2+\rp^2 z'_{\pm}\right)\bigg\}\nonumber\\
~&& +\frac{\Gamma (a'_{\pm}) \delta ^{b'_{\pm}}}{\Gamma (a'_{\pm}-b'_{\pm}+1)} \bigg\{\, _1\tilde{F}_1(a'_{\pm};b'_{\pm};z'_{\pm} \delta ) \left(\delta  \rm^2 (a'_{\pm}-b'_{\pm})+z'_{\pm} (\rp-\delta  \rm)^2\right)\nonumber\\
~&& +\delta  \rm (a'_{\pm}-b'_{\pm}) \, _1\tilde{F}_1(a'_{\pm};b'_{\pm}+1;z'_{\pm} \delta ) (-b'_{\pm} \rm+\delta  \rm z'_{\pm}-2 \rp z'_{\pm})\bigg\}\Bigg]~,
\end{eqnarray}
where $ _1\tilde{F}_1(a;b;z)$ denotes the regularized confluent hypergeometric function, which is related to the confluent hypergeometric function by $ _1\tilde{F}_1(a;b;z)\,=\, _1 F_1(a;b;z)/\Gamma(b)$ and $\Gamma(b)$ denotes the Gamma function.

\section{Evaluation of $\mathcal{I}_{j}$ and $\mathcal{I}_{\varepsilon}$ in an extremal \rn spacetime}\label{Appn:Ij-Ie-E}

\subsection{Evaluation of $\mathcal{I}_{A_{\omega_{k}}}$}\label{Appn:Ij-E}

For the extremal case, from Eq. (\ref{eq:IA-RN-E-2}) let us consider that $\mathcal{I}_{A_{\omega_{k}}} = ((\rs/2)^2/4\pi\omega_{k})\,|\bar{\mathcal{I}}_{A_{\omega_{k}}}|^2$. The integral $\bar{\mathcal{I}}_{A_{\omega_{k}}}$ can be evaluated analytically introducing a regulator of the form $e^{-\epsilon \,\bar{y}_{A}-\epsilon /\bar{y}_{A}}\,\bar{y}_{A}^{\epsilon}$, with small positive and real $\epsilon$. To provide a methodical and straightforward estimation of the integral $\bar{\mathcal{I}}_{A_{\omega_{k}}}$, let us first recall some formulas like
\begin{eqnarray}\label{eq:AppnC-E-1}
\int_{0}^{\infty}d\bar{x}\, \bar{x}^{\bar{z}-n}\, e^{-\left(\bar{a}\, \bar{x}+\frac{\bar{b}}{\bar{x}}\right)} = 2\, \bar{a}^{\frac{1}{2} (n-\bar{z}-1)} \bar{b}^{\frac{1}{2} (-n+\bar{z}+1)} K_{n-\bar{z}-1}\left(2 \sqrt{\bar{a}\,\bar{b}}\right)~,
\end{eqnarray}
where to obtain this formula the conditions $Real[\bar{a}]>0$ and $Real[\bar{b}]>0$ needs to be satisfied (for reference see page $337$ of \cite{gradshteyn2007}). Here the function $K_{n}\left(z\right)$ denotes the modified Bessel function of second kind of order $n$. In particular, the explicit representation of $\bar{\mathcal{I}}_{A_{\omega_{k}}}$ from (\ref{eq:IA-RN-E-2}) is
\begin{eqnarray}\label{eq:AppnC-E-2}
\bar{\mathcal{I}}_{A_{\omega_{k}}} &=& 2\,e^{-i \rs (\Delta E+\omega_{k})}\int_{0}^{\infty} d\bar{y}_{A}\, (\bar{y}_{A}+1)^2 ~ \bar{y}_{A}^{-2+2 i \rs (\Delta E+\omega_{k})} \nonumber\\
 ~&&~\times~ \exp \left(\frac{1}{2} i \rs \bar{y}_{A} (\Delta E+2 \omega_{k})-\tfrac{i \rs  (\Delta E+\omega_{k})}{\bar{y}_{A}}\right)~.
\end{eqnarray}
In this expression we introduce the regulator $e^{-\epsilon \,\bar{y}_{A}-\epsilon /\bar{y}_{A}}\,\bar{y}_{A}^{\epsilon}$ and consider a redefinition of parameters $\bar{x}=\bar{y}_{A}$, $\bar{a}=\epsilon - i \rs (\Delta E+2\omega_{k})/2$, $\bar{b}=\epsilon+i \rs  (\Delta E+\omega_{k})$, and $\bar{z}=\epsilon+2 i \rs (\Delta E+\omega_{k})$, which also satisfies the condition $Real[\bar{a}]>0$ and $Real[\bar{b}]>0$. Then the previous integral can be expressed as
\begin{eqnarray}\label{eq:AppnC-E-3}
\bar{\mathcal{I}}_{A_{\omega_{k}}} &=& 2\,e^{-i \rs (\Delta E+\omega_{k})}  \int_{0}^{\infty}d\bar{x}\,(\bar{x}+1)^2\, \bar{x}^{\bar{z}-2}\, e^{-\left(\bar{a}\, \bar{x}+\frac{\bar{b}}{\bar{x}}\right)}~.
\end{eqnarray}
The evaluation of this integral is now straightforward using the formula from Eq. (\ref{eq:AppnC-E-1}), and results in 
\begin{eqnarray}\label{eq:AppnC-E-4}
\bar{\mathcal{I}}_{A_{\omega_{k}}} &=& 4\,e^{-i \rs (\Delta E+\omega_{k})}  \bar{a}^{-\frac{\bar{z}}{2}}\, \bar{b}^{\bar{z}/2} \Bigg[\left(2-\frac{\bar{z}}{\bar{b}}\right) K_{\bar{z}}\left(2 \sqrt{\bar{a}\,\bar{b}} \right)+\frac{(\bar{a}+\bar{b}) K_{\bar{z}+1}\left(2 \sqrt{\bar{a}\,\bar{b}}\right)}{\sqrt{\bar{a}\,\bar{b}}}\Bigg]~.
\end{eqnarray}

\subsection{Evaluation of $\mathcal{I}_{\varepsilon_{\omega_{k}}}$}\label{Appn:Ie-E}

Next to evaluate $\bar{\mathcal{I}}^{W}_{\varepsilon}(\omega_{k})$ from Eq. (\ref{eq:IeW-RN-E-1}) in the extremal case we express the integral as
\begin{eqnarray}\label{eq:AppnC-E-IeW-1}
\bar{\mathcal{I}}^{W}_{\varepsilon}(\omega_{k}) &=& \frac{\rs^2}{2\,\omega_{k}}\,e^{i\omega_{k} \,(d_{n}-c_{1}-d_{s})}\,e^{i\Delta E \,d_{n}}~e^{-i \rs \Delta E/2}\nonumber\\
~&& \times~\int_{0}^{\infty} d\bar{y}_{A}\, (\bar{y}_{A}+1)^2 ~ \bar{y}_{A}^{-2+2 i \rs (\Delta E+\omega_{k})} \nonumber\\
 ~&&~\times~ \exp \left(\frac{1}{2} i \rs \bar{y}_{A} (\Delta E+2 \omega_{k})-\tfrac{i \rs  (\Delta E+\omega_{k})}{\bar{y}_{A}}\right)~.
\end{eqnarray}
Introducing regulator of the form $e^{-\epsilon \,\bar{y}_{A}-\epsilon /\bar{y}_{A}}\,\bar{y}_{A}^{\epsilon}$ and with redefined parameters $\bar{x}=\bar{y}_{A}$, $\bar{a}=\epsilon - i \rs (\Delta E+2\omega_{k})/2$, $\bar{b}=\epsilon+i \rs  (\Delta E+\omega_{k})$, and $\bar{z}=\epsilon+2 i \rs (\Delta E+\omega_{k})$ this previous integral becomes
\begin{eqnarray}\label{eq:AppnC-E-IeW-2}
\bar{\mathcal{I}}^{W}_{\varepsilon}(\omega_{k}) &=& \frac{\rs^2}{2\,\omega_{k}}\,e^{i\omega_{k} \,(d_{n}-c_{1}-d_{s})}\,e^{i\Delta E \,d_{n}}~e^{-i \rs \Delta E/2} \times\int_{0}^{\infty}d\bar{x}\,(\bar{x}+1)^2\, \bar{x}^{\bar{z}-2}\, e^{-\left(\bar{a}\, \bar{x}+\frac{\bar{b}}{\bar{x}}\right)}~\nonumber\\
~&=& \frac{\rs^2}{4\,\omega_{k}}\,e^{i\omega_{k} \,(d_{n}-c_{1}-d_{s})}\,e^{i\Delta E \,d_{n}}\,e^{i \rs (\Delta E+2\omega_{k})/2} ~\bar{\mathcal{I}}_{A_{\omega_{k}}}.
\end{eqnarray}
Using the expression of $\bar{\mathcal{I}}_{A_{\omega_{k}}}$ from Eq. (\ref{eq:AppnC-E-4}) one can easily find out the entire expression of $\bar{\mathcal{I}}^{W}_{\varepsilon}(\omega_{k})$. One should note that this situation is methodically similar to the nonextremal case.

To evaluate $\bar{\mathcal{I}}^{R}_{\varepsilon}(\omega_{k})$ from Eq. (\ref{eq:IeR-RN-E-1}) corresponding to the extremal scenario we express $\bar{\mathcal{I}}^{R}_{\varepsilon}(\omega_{k}) = \bar{\mathcal{I}}^{R^{+}}_{\varepsilon}(\omega_{k})+\bar{\mathcal{I}}^{R^{-}}_{\varepsilon}(\omega_{k})$. Here $\bar{\mathcal{I}}^{R^{\pm}}_{\varepsilon}(\omega_{k})$ are given by
\begin{eqnarray}\label{eq:AppnC-E-IeR-1}
 \bar{\mathcal{I}}^{R^{\pm}}_{\varepsilon}(\omega_{k}) &=& -\frac{\rs^2}{4\,\omega_{k}}\,e^{i\Delta E(d_{n}-c_{1})} e^{\pm i \,d_{s} \,\omega_{k}} e^{ -i \rs\Delta E} \nonumber\\
 &\times& \int_{0}^{\infty} d\bar{y}_{A}\, 
\, (\bar{y}_{A}+1)^2 \,\bar{y}_{A}^{-2+i \rs (3 \Delta E\mp \omega_{k})} ~\exp \left[i \rs \left\{\bar{y}_{A}(2\Delta E\mp\omega_{k})-\frac{(3\Delta E\mp\omega_{k})}{\bar{y}_{A}}\right\} \right] .
\end{eqnarray}
We introduce regulators of the form $e^{-\epsilon \,\bar{y}_{A}-\epsilon /\bar{y}_{A}}\,\bar{y}_{A}^{\epsilon}$ and redefine parameters as $\bar{x}=\bar{y}_{A}$, $\bar{a}'_{\pm}=\epsilon - i \rs (2\Delta E\mp\omega_{k})$, $\bar{b}'_{\pm}=\epsilon+i \rs  (3\Delta E\mp\omega_{k})$, and $\bar{z}'_{\pm}=\epsilon+ i \rs (3\Delta E\mp\omega_{k})$. Then using the integral representation from Eq. (\ref{eq:AppnC-E-1}) one can find out the expressions of the concerned integrals as
\begin{eqnarray}\label{eq:AppnC-E-IeR-2}
 \bar{\mathcal{I}}^{R^{\pm}}_{\varepsilon}(\omega_{k}) &=& -\frac{\rs^2}{2\,\omega_{k}}\,e^{i\Delta E(d_{n}-c_{1})} e^{\pm i \,d_{s} \,\omega_{k}} e^{ -i \rs\Delta E} \nonumber\\
 &\times&   \bar{a}^{-\bar{z}'_{\pm}/2}~ \bar{b}_{\pm}^{\prime\,\bar{z}'_{\pm}/2} \Bigg[\left(2-\frac{\bar{z}'_{\pm}}{\bar{b}'_{\pm}}\right) K_{\bar{z}'_{\pm}}\left(2 \sqrt{\bar{a}'_{\pm}\,\bar{b}'_{\pm}} \right)+\frac{(\bar{a}'_{\pm}+\bar{b}'_{\pm}) K_{\bar{z}'_{\pm}+1}\left(2 \sqrt{\bar{a}'_{\pm}\,\bar{b}'_{\pm}}\right)}{\sqrt{\bar{a}'_{\pm}\,\bar{b}'_{\pm}}}\Bigg]~.
\end{eqnarray}
Here $K_{n}\left(z\right)$ denotes the modified Bessel function of second kind of order $n$.

\end{widetext}

\end{appendix}

\bibliographystyle{apsrev}
\bibliography{bibtexfile}

\end{document}